\documentclass[fleqn,usenatbib]{mnras}

\usepackage{newtxtext,newtxmath}

\usepackage[T1]{fontenc}

\DeclareRobustCommand{\VAN}[3]{#2}
\let\VANthebibliography\thebibliography
\def\thebibliography{\DeclareRobustCommand{\VAN}[3]{##3}\VANthebibliography}

\usepackage{graphicx}	
\usepackage{amsmath}	

\usepackage{amssymb}	

\title[A RLOF sdO + WD binary: a CE detected]{A Roche Lobe-filling hot Subdwarf and White Dwarf Binary: possible detection of an ejected common envelope}

\author[Jiangdan Li et al.]
{Jiangdan Li$^{1,2,3}$\thanks{E-mail: lijiangdan@ynao.ac.cn},
Christopher A. Onken$^{4,5}$,
Christian Wolf$^{4,5}$\thanks{E-mail: christian.wolf@anu.edu.au},
P\'eter N\'emeth$^{6,7}$,
Mike Bessell$^{4}$,
\newauthor Zhenwei Li$^{1,2}$,
Xiaobin Zhang$^{8}$,
Jiao Li$^{9}$,
Luqian Wang$^{1}$,
Lifang Li$^{1,2}$,
Yangping Luo$^{10}$,
\newauthor Hailiang Chen$^{1,2}$,
Kaifan Ji$^{1}$,
Xuefei Chen$^{1,2}$,
Zhanwen Han$^{1,2,3}$\thanks{E-mail: zhanwenhan@ynao.ac.cn}
\\
$^{1}$Yunnan Observatories, Chinese Academy of Sciences (CAS), 396 Yangfangwang, Guandu District, Kunming 650011, P.R. China\\
$^{2}$Key Laboratory for the Structure and Evolution of Celestial Objects, CAS, Kunming 650011, P.R. China\\
$^{3}$University of Chinese Academy of Sciences, Beijing 100049, China\\
$^{4}$Research School of Astronomy and Astrophysics, Australian National University, Weston Creek ACT 2611, Australia\\
$^{5}$Centre for Gravitational Astrophysics, Australian National University, Canberra ACT 2600, Australia\\
$^{6}$Astronomical Institute of the Czech Academy of Sciences, CZ-25\,165, Ond\v{r}ejov, Czech Republic\\
$^{7}$Astroserver.org, F\H{o} t\'er 1, 8533 Malomsok, Hungary\\
$^{8}$Key Laboratory of Optical                  Astronomy, National Astronomical Observatories, CAS, Beijing, 100012, China\\
$^{9}$Key Lab of Space Astronomy and Technology, National Astronomical Observatories, Chinese Academy of Sciences, Beijing 100101, China\\
$^{10}$Department of Astronomy, China West Normal University, Nanchong, 637002, PR China\\
}
\date{Accepted XXX. Received YYY; in original form ZZZ}

\pubyear{2022}

\begin{document}
\label{firstpage}
\pagerange{\pageref{firstpage}--\pageref{lastpage}}
\maketitle

\begin{abstract}
Binaries consisting of a hot subdwarf star and an accreting white dwarf (WD) 
are sources of gravitational wave radiation at low frequencies and possible progenitors of type Ia supernovae if the WD mass is large enough. 
Here, we report the discovery of the third binary known of this kind:
it consists of a hot subdwarf O (sdO) star and a WD with an orbital period of 3.495 hours and an orbital shrinkage of 0.1\,s in 6 yr. The sdO star overfills its Roche lobe and likely transfers mass to the WD via an accretion disk.
From spectroscopy, we obtain an effective temperature of 
$T_{\mathrm{eff}}=54\,240\pm1\,840$\,K and a surface gravity of 
$\log{g}=4.841\pm0.108$ for the sdO star.
From the light curve analysis, we obtain a sdO mass of $M_{\mathrm{sdO}}=0.55$\,${\mathrm{M_{\odot}}}$ and a mass ratio of $q=M_{\mathrm{WD}}/M_{\mathrm{sdO}}=0.738\pm0.001$. Also, we estimate that the disk has a radius of $\sim 0.41R_\odot$ and a thickness of $\sim 0.18R_\odot$.
The origin of this binary is probably a common envelope ejection channel, where the progenitor of the sdO star is either an RGB star or, more likely, an early AGB star; the sdO star will subsequently evolve into a WD and merge with its WD companion, likely resulting in an R CrB star.
The outstanding feature in the spectrum of this object is strong Ca~H\&K lines, which are blueshifted by $\sim$200\,km/s and likely originate from the recently ejected common envelope, and we estimated that the remnant CE material in the binary system has a density $\sim 6\times 10^{-10} {\rm g/cm^3}$.

\end{abstract}




\section{Introduction}
\label{sec:Introduction}
{\sc small capital letters}
Hot subdwarf O/B (sdO/B) stars have similar surface temperatures and spectral features as main-sequence O/B stars but have lower luminosities of only 10 to 100\,$L_\odot$; with effective temperatures up to 55\,000\,K, their $\log {g}$ ranges from 5.1 to 6.4 \citep[in cgs units,][]{Heber2009, Heber2016}.
They are mostly thought to be compact He-burning stars with masses of $\sim$0.5\,${\mathrm{M_{\odot}}}$ and a thin hydrogen envelope surrounding the core \citep{Heber1986, Heber2009, Heber2016}. They are probes in studies of asteroseismology (e.g. \citealt{Charpinet1996, Fontaine2003}) and the structure of the Galaxy \citep{Green1986}. Studies of sdO/B stars help us to understand the evolution and population of binaries \citep{Han2002, Han2003, Han2020}. 

Hot subdwarf binaries are proposed as one of several possible progenitors of type Ia supernovae \citep{Wang2009, Wang2010, Geier2013, Geier2015} and as sources of gravitational waves \citep{Wu2018, Wu2020}. sdO/B + WD binaries with accreting WDs have two possible destinies \citep{Brooks2015, Neunteufel2016, Bauer2017, Brooks2017b, Schwab2019, Schwab2021a, Schwab2021b, Bauer2021, Wong2021}: in one case, the accreted helium layer of the WD is ignited, burns explosively, and may disrupt the WD star, resulting in a double-detonation type Ia supernova \citep{Livne1990, Livne1995, Woosley2011, Wang2012, Shen2014, Neunteufel2017, Wang2018}; this could happen even if the mass of WD is below the Chandrasekhar limit. In the other case, the helium layer on the surface of the WD burns steadily and increases the CO mass of the WD until it exceeds the Chandrasekhar mass and the CO core detonates in a classical type Ia supernova \citep{Bildsten2007, Brooks2015}. In addition to the classical doubledetonation scenario and the Chandrasekhar massscenarios, we found in \cite{Ruiter2013} that such systems (where a CO WD accretes from a stripped He-burning star - i.e. a hot subdwarf) can eventually, after $\sim$ Gyr, lead to a double WD merger.

A large fraction of sdO/B stars is observed to be in binary systems. Most of them have short orbital periods of 
$<$10\,d \citep{Maxted2001, Napiwotzki2004}. The most compact system has a period of less than 1 hour (e.g. \citealt{Vennes2012, Geier2013, Kupfer2017a, Kupfer2017b}). The short periods are likely caused by the loss of orbital angular momentum due to the radiation of gravitational waves after the binaries have survived a common-envelope phase \citep{Han2002, Han2003, Nelemans2010}. 
Only six known sdB binaries are in tight systems with a period of less than 2 hours; these will likely experience Roche-lobe overflow before the sdB evolves into a WD \citep{Vennes2012, Geier2013, Kupfer2017a, Kupfer2017b, Pelisoli2021}. Of these, only two subdwarfs are Roche-lobe filling and in accreting binaries, ZTF J2130+4420 and ZTF J2055+4651 \citep{Kupfer2020a, Kupfer2020b}.

In this paper, we present the discovery of the sdO + WD binary SMSS J192054.50-200135.5 (hereafter J1920-2001).
The Roche-lobe filling sdO has a temperature of $T_{\mathrm{eff}}=54\,240\pm1\,840$\,K, and the WD is accreting from the sdO star. We discovered this object in the SkyMapper Southern Survey\footnote{\url{https://skymapper.anu.edu.au}} \citep[SMSS,][]{Wolf2018, SKYMAPPER2019}, serendipitously in 2019. In Section~\ref{sec:Observations}, we present photometric and spectroscopic data. In Section~\ref{sec:Atmospheric}, we derive the atmospheric parameters from the spectra. In Section~\ref{sec:Light}, we fit the light curve for the physical and orbital parameters of the binary. In Section~\ref{sec:parallax} we discuss the spectral energy distribution and the $U-V$ velocity diagram, and in Section~\ref{sec:formation} the formation channels and destiny. This is followed by a discussion of Ca~II lines in Section~\ref{sec:CaII} and a summary in Section~\ref{sec:summary}.

\section{Observations}
\label{sec:Observations}
We first identified J1920-2001 during a search for short-term variable blue stars in the Data Release 2 (DR2) of the SkyMapper Southern Survey \citep[SMSS,][]{SKYMAPPER2019}. Part of the SMSS observations involve visiting a sky position three times within 20 minutes in the SkyMapper $u$-band, which is narrower and bluer than the eponymous SDSS filter, with a central wavelength and FWHM of 349 and 42\,nm, respectively. 

In order to confirm variability periods of selected objects, these were followed up in Discretionary Time with dedicated SkyMapper pointings. Between 25 June 2019 and 23 October 2019, J1920-2001 was successfully observed on 13 nights with at least 30 exposures in each of two filters, using a $u-i-u-i$ pattern alternating between 100-second exposures in the $u$-band and 40-second exposures in the $i$-band. Together with sparser data from other nights within the above period, over 700 exposures were taken in each filter of which more than 500 were of suitable quality. From this data, the $\sim$3.5-hour period in the light curve of J1920-2001 was first identified.

We then realised that the object was in the footprint of one of the fields of the NASA {\it Kepler} {\it K2} mission. {\it Kepler} was launched in 2009 and has observed approximately 200\,000 objects \citep{Borucki2011, Batalha2013}. The {\it K2} mission makes use of the {\it Kepler} spacecraft after the loss of a reaction wheel and started operations in May 2014 \citep{Howell2014}. Photometric data of J1920-2001 were obtained by {\it Kepler} over 81 days in Campaign 7 of the {\it K2} mission, which lasted from 4 Oct to 26 Dec 2015. There are 3\,432 observations in total, and each observation took a 30-minutes exposure. We used the reduced presearch data conditioning simple aperture photometry (PDCSAP) flux from {\it K2}, which is already systematically corrected by the reduction pipeline for every trend of non-astrophysical origin.  We normalized the light curve using the Lightkurve version 2.0 \citep{LightkurveCollaboration2018}. Lightkurve is a package based on Python that is designed for light curve analysis, especially for {\it Kepler} data. 

Additionally, J1920-2001 was observed with the Palomar Samuel Oschin 48-inch (P48) telescope of the Zwicky Transient Facility \citep[ZTF,][]{Bellm2019b, Bellm2019a, Graham2019}. The high-cadence data covers observations from 2018 to 2021, where each visit has a 30-second exposure \citep{Masci2019}. We selected high-quality data with catflags of zero. In the ZTF-$g$ band, there are 121 observations in total with 19, 33, 27, and 39 visits in 2018, 2019, 2020, and 2021, respectively. In the ZTF-$r$ band, there are 387 observations in total with 22, 305, 27, and 50 visits in the same periods. This number of visits allows finding a period for each year separately. 

Spectroscopic observations of J1920-2001 were carried out with the Wide Field Spectrograph \citep[WiFeS;][]{Dopita2007, Dopita2010} on the Australian National University (ANU) 2.3m-telescope at Siding Spring Observatory. WiFeS has been operational since May 2009 and is currently the main instrument at this telescope. We use spectra taken at a resolution of R=$\lambda/\Delta \lambda \approx 7\,000$, covering a wavelength range from 3\,750\,{\AA} to 4\,370\,\AA. First spectra to type the object were taken on 2019-06-28 as 10-minutes exposures, and spectral time series covering at least part of a period were obtained on 2019-10-22 and 2021-08-13 using 20-minute exposures. 

The spectral data were reduced using the Python-based pipeline PyWiFeS \citep{Childress2014}. PyWiFeS calibrates the raw data with bias, arc, wire, internal-flat, and sky-flat frames, and performs flux calibration and telluric correction with standard star spectra. Flux densities were calibrated using several standard stars throughout the year, which are usually observed on the same night. We then extracted spectra from the calibrated three-dimensional (3D) cube using a bespoke algorithm for fitting and subtracting the sky background after masking sources in a white-light stack of the 3D cube. A few of the spectra were reduced in a similar fashion using FIGARO\footnote{\url{https://starlink.eao.hawaii.edu/docs/sun86.pdf}} \citep{Shortridge92, Shortridge04}. Afterwards, we select reliable spectra with a signal-to-noise ratio (SNR) larger than 5, and perform a normalization process using the function in ISpec \citep{Blanco-Cuaresma2014a, Blanco-Cuaresma2019}. One of the observed WiFeS spectra is shown in Figure~\ref{fig:spectra_fits} as an example.

\section{Atmospheric parameters}
\label{sec:Atmospheric}
\subsection{Spectral fitting}
\label{subsec:Spectra}
\begin{figure}

	\includegraphics[width=\columnwidth]{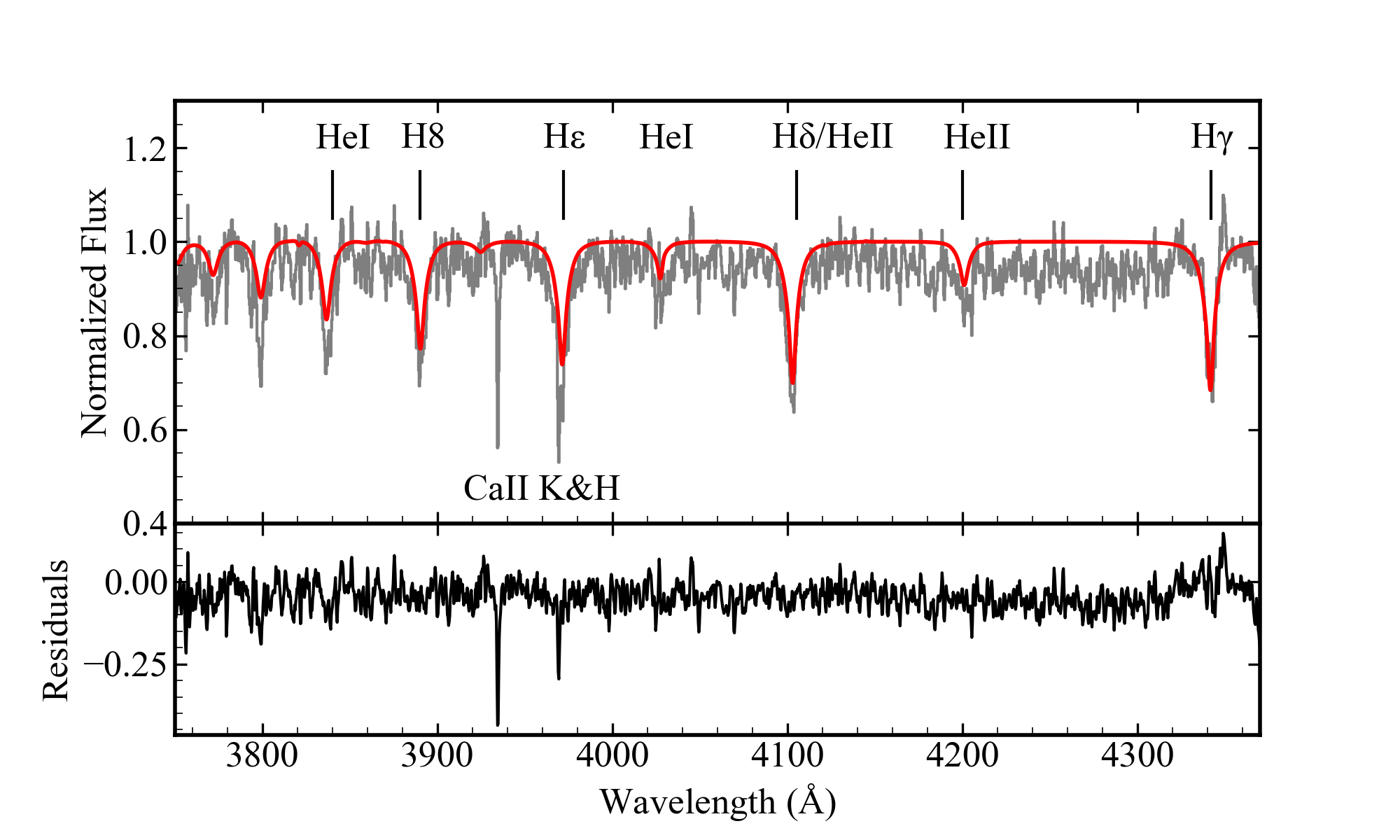}
    \caption{Top: the observed spectrum from 22 Oct 2019 (grey line) and the best-fit model (red line). The main absorption lines are labelled, including Balmer lines, Helium lines and the  
    Ca~II H\&K lines. The parameters of the best fit are $T_{\mathrm{eff}}=54\,240\pm1\,840$\,K, $\log{g}=4.841\pm0.108$, $\log(n({\mathrm{He}})/n({\mathrm{H}}))=-1.190\pm0.142$ and $v_{\mathrm{rot}}{\mathrm{sin}}i=143.47\pm7.5$\,km/s.
    Bottom: fit residuals. }
    \label{fig:spectra_fits}
\end{figure}

In Figure~\ref{fig:spectra_fits}, we show an example spectrum observed on 2019-10-22. It ranges from 3\,750\,{\AA} to 4\,370\,{\AA}, covering the Balmer lines H8 $\lambda$3\,889.06, H$\varepsilon$ $\lambda$3\,970.08, H$\delta$ $\lambda$4\,101.73 and H$\gamma$ $\lambda$4\,340.47, as well as the neutral He~I $\lambda$3\,833.55, $\lambda$4\,026.19 and ionized He~II $\lambda$4\,102, $\lambda$4\,200 lines in air wavelength. The spectrum is single-lined with the lines of the primary sdO star. 
The Ca~{II}~H $\lambda$3\,968.47 and K $\lambda$3\,933.66 lines have a constant radial velocity and thus do not originate in the stellar atmospheres. 
Atmospheric parameters of effective temperature $T_{\mathrm{eff}}$, surface gravity, $\log {g}$, helium abundance $\log {y}$, where $y=n({\mathrm{He}})/n({\mathrm{H}})$, and projected rotational velocity were all determined for the sdO star by fitting the normalized spectra with NLTE (HHeCNO) atmospheric models from {\sc XTgrid} \citep{Nemeth2019}. 
First, a $\chi^{2}$-minimization was performed to constrain effective temperature, surface gravity and helium abundance, while the rotational velocity was ignored. We find $T_{\mathrm{eff}}=54\,240\pm1\,840$\,K, $\log{g}=4.841\pm0.108$, $\log(n({\mathrm{He}})/n({\mathrm{H}}))=-1.190\pm0.142$. Then the projected rotational velocity was derived for each spectrum individually using the best-fit atmospheric parameters. We find $v_{\mathrm{rot}}{\mathrm{sin}} i=143.47\pm7.5$\,km/s. 
Using two steps in deriving the parameters avoids degeneracies between the surface gravity and the projected rotational velocity.
We show the fit to the WiFeS spectrum in Figure~\ref{fig:spectra_fits}. In the upper panel, the red line is the spectrum of the model sdO star while the grey line is our observed spectrum. The bottom panel shows the residuals between the model and our observation. Except at the position of the Ca~{II}~H\&K lines (see Section~\ref{sec:CaII}), the residuals are consistent with observed noise, which means that the spectrum fits that of a sdO star very well.

\subsection{Radial velocity curve fitting}
\label{subsec:Radial}
The spectra in the time series show different radial velocities (RVs) caused by the binary orbit. We first generate a spectral template with iSpec \citep{Blanco-Cuaresma2014a,Blanco-Cuaresma2019} and the TLUSTY atmospheric models of O-type stars \citep{Lanz2003} to interpolate a spectrum with $T_{\mathrm{eff}}=54\,240$\,K, $\log {g}=4.75$ and $v_{\mathrm{rot}}{\mathrm{sin}}i=143.47$\,km/s. The resolution is set to 7\,000 and the wavelength range to cover 3\,700\,\AA\ to 4\,400\,\AA. To fit the periodic RV variability, we cross-correlate the individual WiFeS spectra with the spectral template and measure velocities using the cross-correlation function (CCF) by iSpec. The CCF between the spectral template and our observed spectrum would show a peak at the radial velocity shift of our spectrum. 

\begin{figure}
	\includegraphics[width=\columnwidth]{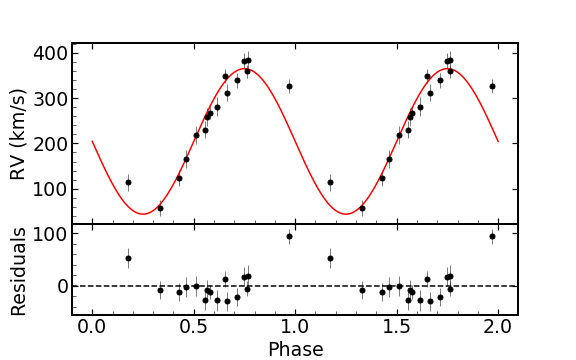}
    \caption{The best fits to our radial velocity curve. In the top panel, the red line is the fitting model, while the black dots are the phased radial velocity. The best-fit parameters of the radial velocity curve are an eccentricity of the binary orbit $e=0$, 
    an amplitude of the primary star $K_1=169.22\pm 2.03$\,km/s, and a systemic velocity of $200.95\pm 4.46$\,km/s. The bottom panel shows the residuals.}
    \label{fig:Rv_curve}
\end{figure}

The radial velocities of the full time series thus produce a radial-velocity curve (see Table~\ref{tab:RV}). 
We use the photometric data from SkyMapper to derive the observational period and $T_0$ in 2019 and data from ZTF for 2021 with 0.145763 and 0.145762\,d (see Section~\ref{subsec:Orbital_period}), respectively, the periods themselves may be consistent between 2019 and 2021, but
the radial velocities are phased differently, with zero points in the ephemeris of 2\,457\,301.1295 and 2\,457\,301.1372\,d in 2019 and 2021, respectively. 
We fit a sine curve to the folded RV data points using the RadVel python package \citep{Fulton2018}. 
Using a Monte-Carlo Markov chain (MCMC), we perform 1\,000 steps to simulate the residuals between the model and the observational data. The resulting orbital parameters of the RV curve fitting show an eccentricity of $e$=0 
a semi-amplitude of the radial-velocity curve of $K_1=169.22\pm 2.03$\,km/s and a systemic velocity of $200.95\pm 4.46$\,km/s.
The red curve in Figure~\ref{fig:Rv_curve} shows our best-fit model with the parameters derived in the MCMC.

\begin{table}
  \centering
  \caption{Measured radial velocity of each observation}
  \setlength{\tabcolsep}{1\tabcolsep}
  \begin{tabular}{@{}ccccc@{}}
\hline\hline
Date & Start time &Exp. time & RV  & Error of RV  \\
     & (bjd)             & (s)          & (km/s)    & (km/s)\\
\hline
2019-06-28 & 2\,458\,645.2688 &600 & 164.98 & 19.39 \\
           & 2\,458\,645.2761 &600   & 218.22 & 19.05\\
           & 2\,458\,645.2834 &600 & 258.11 & 18.92\\
           & 2\,458\,645.2907 &600 & 281.19 & 19.97\\
           & 2\,458\,645.2980 &600 & 311.77 & 18.25\\
           & 2\,458\,645.3053 &600 & 338.68&17.44\\
           & 2\,458\,645.3127  &600 &358.50 & 14.84\\
2019-10-22 & 2\,458\,778.9071  &1\,200&57.02&17.43\\
           & 2\,458\,778.9213  &1\,200&123.66&17.33\\
           & 2\,458\,778.9396  &1\,200&229.84&18.80\\
           & 2\,458\,778.9538  &1\,200&348.50&15.21\\
           & 2\,458\,778.9681  &1\,200&382.25&17.61\\
2021-08-13 & 2\,459\,439.8655  &600&383.28&19.36\\
           & 2\,459\,439.8886  &1\,200&327.04&15.22\\
           & 2\,459\,439.9181  &1\,200&114.31&19.16\\
           & 2\,459\,439.9770  &1\,200&267.64&13.93\\

\hline\hline
\end{tabular}
\label{tab:RV}
\end{table}

\section{Light Curve Modelling}
\label{sec:Light}
\subsection{Orbital period}
\label{subsec:Orbital_period}
\begin{figure}

	\includegraphics[width=\columnwidth]{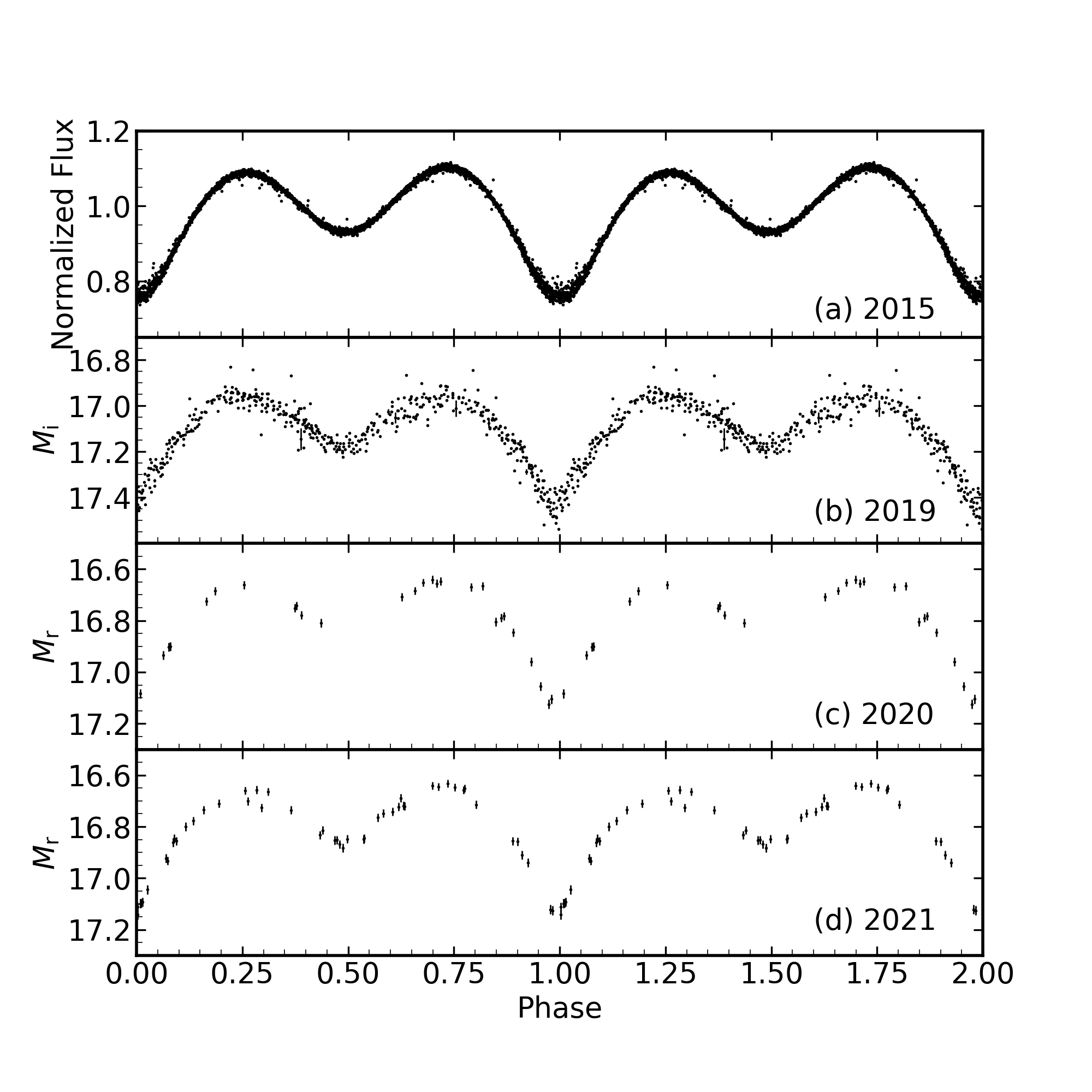}
    \caption{Phased light curves for 2015, 2019, 2020 and 2021, based on data from {\it K2}, SkyMapper, ZTF and ZTF, respectively. The period and the ephemeris zero point are shown in Table~\ref{tab:period}. We show two cycles in the figure,  the sdO is furthest away from the observers and the WD is closest to the observers at phase 0 or 1, the WD is furthest away from the observers and the sdO is closest to the observers at phases 0.5 or 1.5.
    }
    \label{fig:LC}
\end{figure}

\begin{figure}
	\includegraphics[width=\columnwidth]{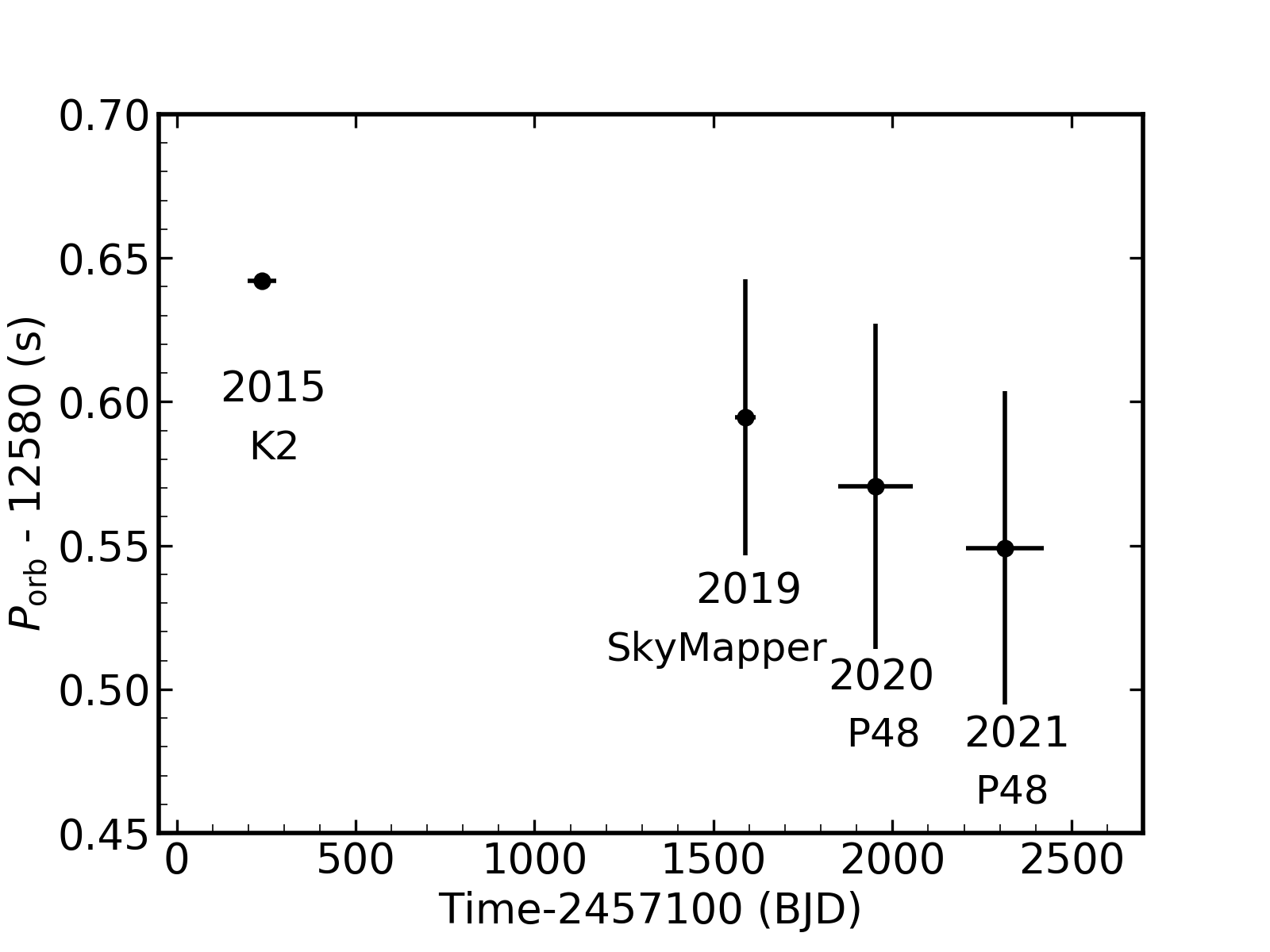}
    \caption{The orbital periods in 2015, 2019, 2020 and 2021. The photometric data we used in 2015, 2019, 2020 and 2021 comes from {\it K2},  SkyMapper, ZTF, and ZTF, respectively. The error-bars of data points along the x-axis are the time interval spanned by the observations each year. Note that the error-bar of the orbital period (the y-axis) for 2015 is small and invisible. The period and the ephemeris zero point are shown in Table~\ref{tab:period}.
    }
    \label{fig:period}
\end{figure}

The photometric data covers five years of observations from {\it K2}, SkyMapper, and ZTF \citep{Howell2014, Masci2019, SKYMAPPER2019} in 2015, 2018, 2019, 2020, and 2021 (see Figure~\ref{fig:LC}). The light curve of J1920-2001 shows strong periodic variability. We used the Lomb Scargle periodogram and an MCMC approach of 5\,000 times progressing to find the period of the light curve for each year \citep{Scargle1982, pya}. 
We do not use the observations in 2018 from ZTF because the small number of observational visits and the quality of the observations do not allow us to obtain a reliable period from the light curve. 
The observational periods of the phased light curves of 2015, 2019, 2020 and 2021 are 0.14576329,  0.14576274, 0.14576246 and 0.14576221\,d, respectively (see Figure~\ref{fig:period}). We chose the deepest point of the phased light curve (i.e. the primary minimum) as the zero point of the ephemeris, which are 2\,457\,301.1279, 2\,457\,301.1295, 2\,457\,301.1326, and 2\,457\,301.1372\,d, respectively.

The spatial motion of the binary system results in a time dilation, which we correct with the Galactic space velocities $(U, V, W)$. 
For the kinematic parameters of J1920-2001 we take a proper motion, 
PM(RA)$=-4.364\pm0.072$ mas\,$\mathrm{yr}^{-1}$ and 
PM(Dec)$=-5.853\pm0.062$ mas\,$\mathrm{yr}^{-1}$, and a parallax of $0.14119\pm 0.03891$ mas from Gaia EDR3 \citep{Gaia_edr3_2021}, and a radial velocity measured from our spectra of 200.95\,km/s.
We compute Galactic space velocities of $U=282.79$\,km/s, $V=86.99$\,km/s and $W=7.14$\,km/s. Relative to the Sun with $U_\odot=11.1$\,km/s, $V_\odot=250.24$\,km/s and $W_\odot=7.25$\,km/s \citep{Schoenrich2010, Schoenrich2012}, the space velocity of J1920-2001 is $v=316.97$\,km/s. 
The true period is shorter than the observed period and can be obtained by multiplying with $1-{v}/{c}=0.9989$, where c is the speed of light and the factor $(1-{v}/{c})$ is the classical value.
After correcting the time dilation, the true periods for 2015, 2019, 2020 and 2021 are 0.14560928, 0.14560873, 0.14560846 and 0.14560821\,d, respectively.  The errors on the periods and the zero points are shown in the brackets of Table~\ref{tab:period}, the errors in the brackets of zero mean the error is smaller than the precision we showed. 
The orbital period of J1920-2001 appears to decrease by $\sim$0.1 seconds from 2015 to 2021.

\begin{table*}
  \centering
  \caption{The orbital period and the zero point of the ephemeris ($T_0$) of 2015, 2019, 2020 and 2021.}
   \setlength{\tabcolsep}{\tabcolsep}
  \begin{tabular}{ccccccc}
\hline\hline
Year & Tele. & $\mathrm{N_{exp}}$ & Exp. time (s) & Observed Period (day) & Real period (day) & $T_0$ (bjd) \\
\hline
2015 & {\it K2}  & 3\,432 &1\,800 & $0.14576329(01)$ & $0.14560928(01)$ & $2\,457\,301.1279(00)$\\
2019 & SkyMapper & 513    & 40    & $0.14576274(55)$ & $0.14560873(55)$ & $2\,457\,301.1295(53)$\\
2020 & P48 (ZTF) & 27     & 30    & $0.14576246(65)$ & $0.14560846(65)$ & $2\,457\,301.1326(77)$\\
2021 & P48 (ZTF) & 50     & 30    & $0.14576221(63)$ & $0.14560821(63)$ & $2\,457\,301.1372(92)$\\

\hline\hline
\end{tabular}
\label{tab:period}
\end{table*}

\subsection{Physical parameters from light curve fitting}
\label{subsec:Orbital_parameters}
We used the {PHysics Of Eclipsing BinariEs (PHOEBE)} code to perform a light curve analysis. PHOEBE is an open-source eclipsing-binary modelling code, which can reproduce and fit light curves \citep{Prsa2016, Prsa2018, Conroy2020}. The {\it K2} light curves are of high precision and have the largest number of data points among the available seasons. We fit the {\it K2} light curves to obtain the physical parameters of the system.

In the light curve modelling, we set the distortion of the stars to the Roche potential and the radiation effect to the Wilson radiation.
We used a semi-detached system to fit the light curve, assuming the sdO star is filling the Roche lobe and transferring matter to the compact secondary.
In the modelling, we have considered the flux emitted from the mesh grid under the influence of limb darkening, gravity darkening, and reflection effects. As the primary star is hotter than $8\,000$\,K and the secondary is a WD, we set the reflection and $\beta$ coefficient for gravity darkening corrections to 1, which means that all incident flux is reflected. The flux has been integrated at each point of the grid by assuming a blackbody of a given temperature at the {\it Kepler} mean bandpass. 
 
In the {\rm K2} light curve fitting, we set the orbit to be a circular orbit with a period of 0.14560928\,d in 2015, as is shown in Section~\ref{subsec:Orbital_period}.  
We also fix the atmospheric parameters of effective temperature ($T_{\mathrm{eff}}$), radial velocity amplitude ($K$), and the surface gravity ($g$) of the sdO star to the values derived from the spectral fitting. 
We adopt the radial velocity amplitude of the primary star $K_1=169.22\pm2.03$\,km/s, derived from our radial velocity curve fitting, and obtain $a_1\sin{i}=0.488\pm0.006\,{\mathrm{R_{\odot}}}$  from the equation of $K_1=\frac{2\pi a_1 \sin{i}}{\sqrt{1-e^2}}$. The value of $a_1\sin{i}$ is used to constrain the mass ratio and the primary mass in the fitting.
 
\begin{figure}
	\includegraphics[width=\columnwidth]{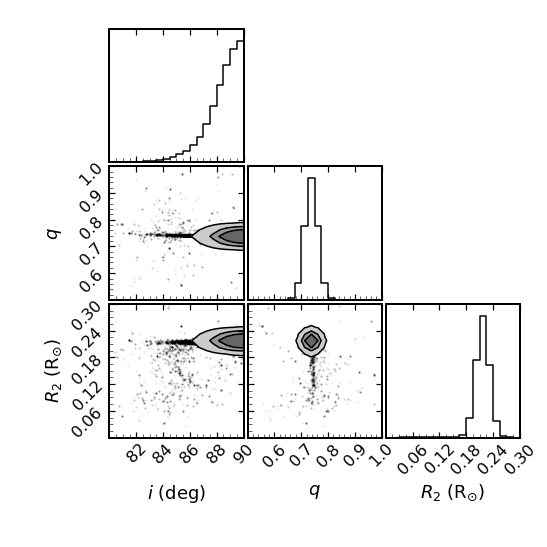}
    \caption{The light curve fitting with PHOEBE results in an orbital inclination of $i=89.61^{+0.27}_{-0.44}$ deg, a mass ratio of $q=M_{\mathrm{WD}}/M_{\mathrm{sdO}}=0.738\pm 0.001$, and a radius of the secondary of $R_{\mathrm{2}}=0.213\pm 0.002\,R_{\odot}$.
    }
    \label{fig:corner}
\end{figure}

We fit our light-curve model to the photometric data from {\it K2}\footnote{The exposure time in {\it K2} is 30 minutes, which induces large smearing in the light curve. Therefore, in our fitting, we need to smear the PHOEBE synthesized light curve by the 30 minutes exposure time, and compare the {\it K2} light curve with the smeared synthesized light curve.}, and the parameters are calculated by MCMC with 10\,000 steps.
We show a corner plot in Figure~\ref{fig:corner} with the resulting parameters of an inclination angle of $89.61^{+0.27}_{-0.44}$ deg, a mass ratio of $q=M_{\mathrm{WD}}/M{\mathrm{sdO}}=0.738 \pm 0.001$ and a radius of the secondary of $R_{\mathrm{2}}=0.213\pm 0.002\,{\mathrm{R_{\odot}}}$. From the physical parameter we derived in previous sections, we have obtained a primary mass of $M_{\mathrm{sdO}}=0.55\,{\mathrm{M_{\odot}}}$, a primary radius of $R_{\mathrm{sdO}}=0.47\,{\mathrm{M_{\odot}}}$ and a secondary mass of $M_{\mathrm{2}}=0.41\,{\mathrm{M_{\odot}}}$. The effective temperature of the secondary is unavailable because the luminosity of the secondary is much less than the primary, and consequently the temperature of the secondary does not change the light curve in the fitting.

\begin{figure}
	\includegraphics[width=\columnwidth]{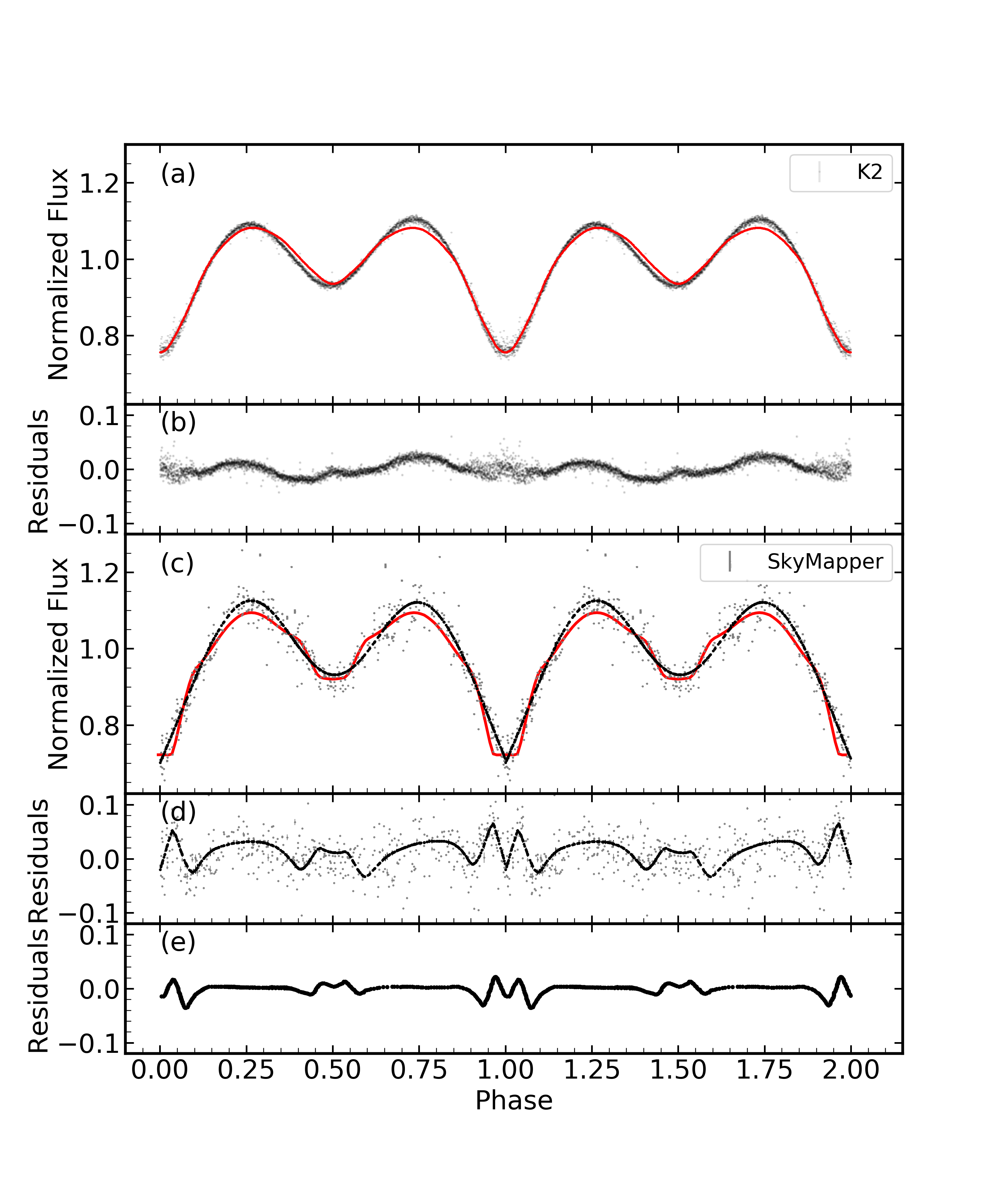}
    \caption{The residuals between the observational and the synthetic light curves. Panel (a) shows the phased light curve from {\it K2} (black dots) and the synthesized PHOEBE light curve with 30-minute smearing (red line). Panel (b) shows the residuals between the {\it K2} observational data and the smeared synthesized light curve. Panel (c) shows the phased light curve from SkyMapper in $i$-band (grey dots), the binned dots (black dots) and the best fitting model using PHOEBE (red line). In panel (d), we show the residuals between the SkyMapper data and the model, where black dots are binned from grey ones.
    Panel (e) shows the residuals of ZTF J2130+4420 fitted by \citet{Kupfer2020a} (see the left panel of Figure 5 in their paper). The residuals at phase 0 or 0.5 are due to the occultation of an accretion disk around the WD. 
    }
    \label{fig:Residuals}
\end{figure}

Using the above physical parameters, we show the synthesized (and smeared with a 30-minute exposure time) light curve and the phased {\it K2} light curve, and the residuals in Figure~\ref{fig:Residuals}. In panel (a), the red line is the smeared synthesized light curve, the grey dots are the phased observational data from {\it K2}. Panel (b) shows the residuals between the observed data and the model.

The SkyMapper light curves are of short exposure time, 40\,s in $i$-band. We did a similar fitting with PHOEBE but without any smearing. Panel (c) of Figure~\ref{fig:Residuals} shows the phased light curve (grey dots) from SkyMapper, the binned light curve with Gaussian process (black dots) \citep{Rasmussen2006},  and the best fitting PHOEBE model (red line). Panel (d) shows the residuals between the SkyMapper data and the model.

From the residuals of Figure~\ref{fig:Residuals}, we see that our two-star PHOEBE model does not fit well at phases 0 and 0.5, presumably due to an occultation of the accretion disk around the WD \citep[see also][]{Kupfer2020a, Kupfer2020b}. \citet{Kupfer2020a, Kupfer2020b} suggest that the characteristic shape of the residuals is the feature of a periodically occulted accretion disk around the WD with mass transferring between the sdO and the WD. We compared the residuals of our light curve (panel(d) of Figure~\ref{fig:Residuals}) to the residuals of the light curve for ZTF J2130+4420 with a two-star model (panel (e) of Figure~\ref{fig:Residuals}), and find them to have a very similar shape. While PHOEBE does not yet have a function to add an accretion disk, we infer 
that J1920-2001 has a similarly occulted accretion disk. As a result, the radius for the secondary of $0.213\pm0.002$\,$R_{\odot}$, formally obtained in the light curve fitting, includes light from both the disk and the WD. The radius of the WD itself is expected to be much smaller. 

\begin{figure}
	\includegraphics[width=\columnwidth]{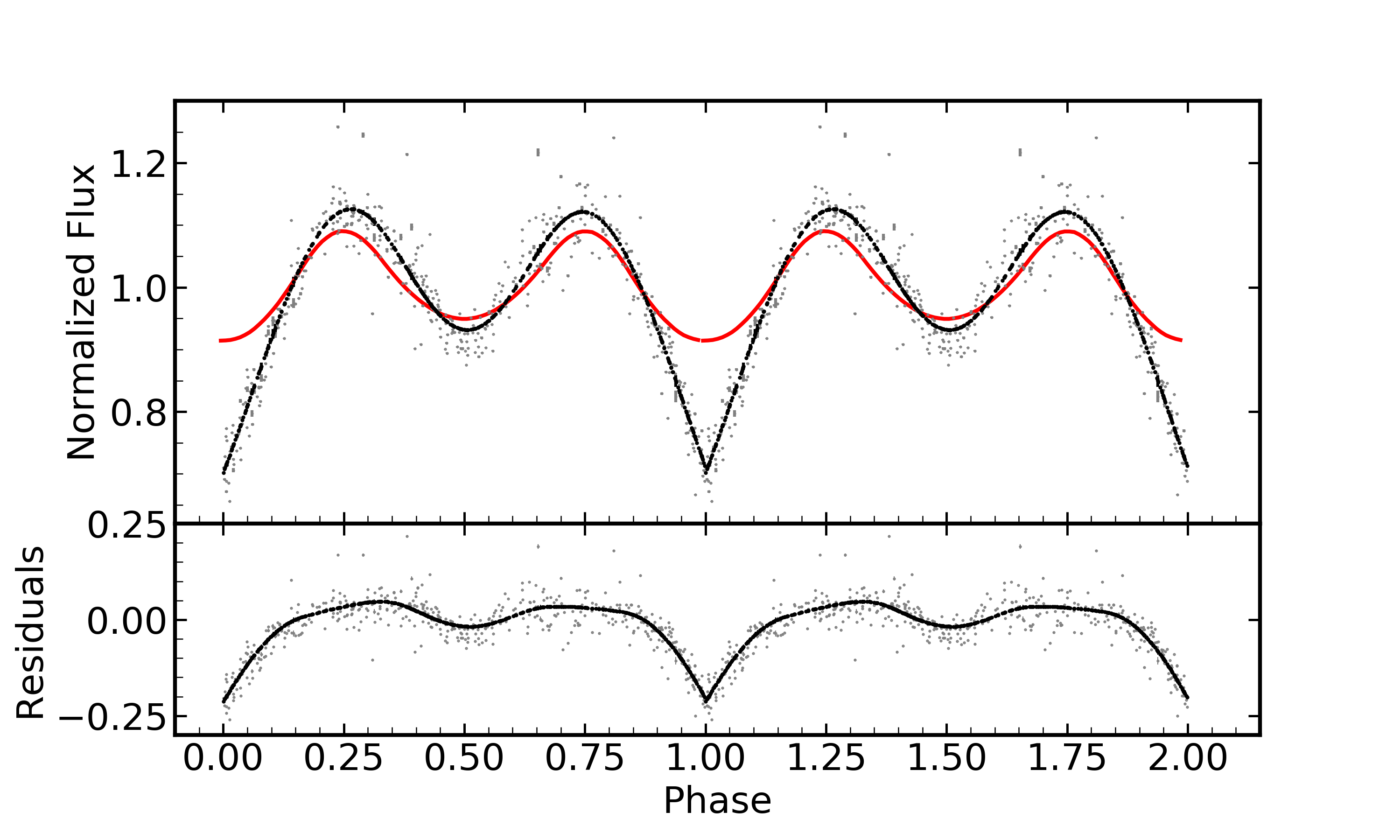}
    \caption{PHOEBE fitting of the SkyMapper light curve.
    The top panel shows the phased light curve from SkyMapper (grey dots), the black dots are the binned data, and the red curve is the PHOEBE two-star fitting model with a sdO and a WD of $R_{\rm WD}=0.0158\,R_{\odot}$. The second panel shows the residuals between the observational data and the fitted model. 
    }
    \label{fig:LC_WD}
\end{figure}

Since the mass of the disk is much less than that of the WD, we assume that the mass of the secondary is that of the WD, i.e. $M_{\rm WD}=0.41\,M_{\odot}$. According to the relation between the WD radius and the WD mass \citep[][]{Nauenberg1972} (see also equation 91 of \cite{Hurley2000}), the calculated radius of the WD is 0.0158\,$R_{\odot}$. Assuming a WD radius of 0.0158\,$R_{\odot}$, and adopting all the other physical parameters we obtained, we generated the synthesized light curve with PHOEBE, which does not have a disk included. The red curve in Figure~\ref{fig:LC_WD} is the synthesized light curve and the residuals are caused by the disk.

The SkyMapper light curve in Figure~\ref{fig:LC_WD} shows a large variability with an inclination of $89.61$\degr. For the edge-on orbit, the primary eclipse occurs when the sdO is furthest away from us,
and the primary minimum should be a flat one if the size of the secondary is much smaller or larger than the primary. However,  Figure~\ref{fig:LC_WD} shows a sharp dip but not a flat minimum for the primary eclipse,
which means that the size of the secondary is likely similar to the primary ($0.47\,R_{\odot}$). 
Also, the spectra show the lines of the primary without any signature of the secondary, which indicates the luminosity of the secondary is much less than the primary.
If the secondary is spherical, e.g. a star, we would see that the luminosity would drop down to nearly zero when the secondary with a size similar to that of the sdO is eclipsing the sdO primary. Obviously, this is not the case here.  Therefore, we show here again that the secondary is a star surrounded by a disk. The disk blocks the luminosity from the sdO and has a size of $\sim 0.47\,R_{\odot}$.

\begin{figure}
	\includegraphics[width=\columnwidth]{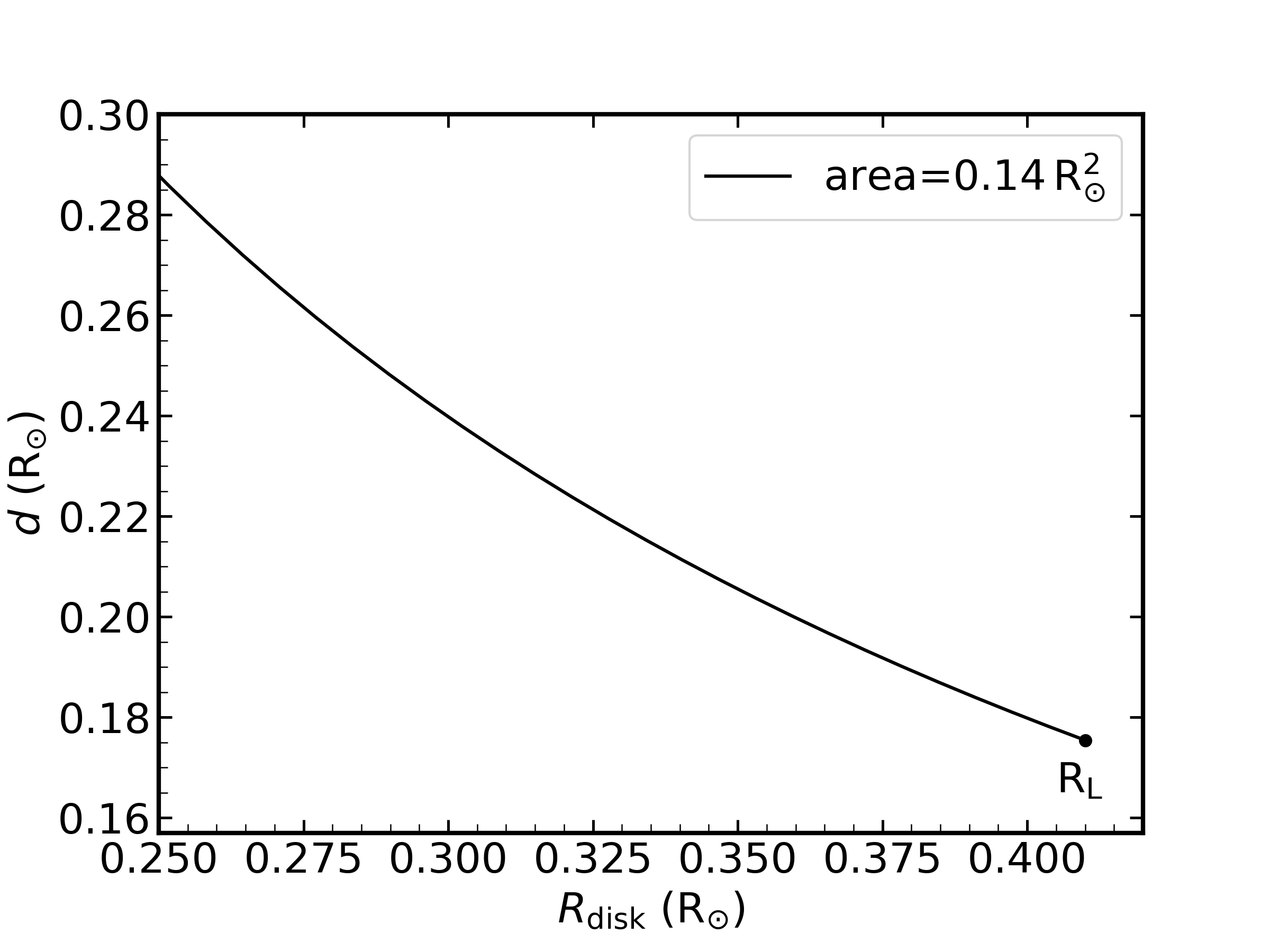}
    \caption{The relation between the radius of the disk, $R_{\rm disk}$, and the thickness of the disk, $d$, which has a blocking area of $0.14\,R_{\odot}^2$. $0.41\,R_{\odot}$ is the radius of the Roche-lobe of the secondary.
    }
    \label{fig:W_Rrelation}
\end{figure}

We try to constrain the disk parameters from the residuals of Figure~\ref{fig:LC_WD}. The temperature of the disk is much less than that of the sdO, and the residuals at the primary minima in Figure~\ref{fig:LC_WD} indicate that the 
disk blocks the luminosity from the sdO with a blocking area of $\sim 0.14\,R_{\odot}^2$, i.e. $\sim 20\%$ of the area of the sdO star.
We assume that the disk has a radius of $R_{\rm disk}$ and a thickness of $d$, and Figure~\ref{fig:W_Rrelation} shows the relation between $R_{\rm disk}$ and $d$ of the disk with a blocking area of $\sim 0.14\,R_{\odot}^2$ with an inclination of $89.61$\degr. The maximum radius of the disk is set to be  $0.41\,{R_\odot}$, which is the radius of the Roche-lobe of the secondary.
The sharp primary minima of the SkyMapper light curve means that the disk is of a similar size to the primary (with a radius of $0.47\,R_{\odot}$), we, therefore, take the radius of the disk to be $R_{\rm disk}=0.41$\,$R_{\odot}$, which is filling the Roche-lobe of the secondary, and consequently the thickness of the disk to be $d=0.18$\,$R_{\odot}$, as shown in Figure~\ref{fig:W_Rrelation}. 

Assuming the sdO star is synchronized to the orbit, 
we obtain the projected rotational velocity of the sdO star, $v_{\mathrm{rot}} \sin{i}=2\pi R_{\mathrm{sdO}}\sin{i}/P_{\rm{orb}}  = 160.82$\,km/s, 
close to the value of $143.47\pm 7.5$\,km/s fitted from the spectra. 

\section {Derived parameters from parallax}
\label{sec:parallax}

\subsection{Fitting the Spectral Energy Distribution}
\label{sec:sed}
From Gaia eDR3, we found a parallax for J1920-2001 of $0.14119\pm 0.03891$ mas, corresponding to a distance of $7\,082 \pm 1\,950$ pc, and a photometric $G$-band magnitude of $m_{\mathrm{G}}=16.53\pm0.009$ mag \citep{GaiaCollaboration2016, GaiaCollaboration2018, Gaia_edr3_2021, Riess2021}. Based on a distance of about 7 kpc and Galactic coordinates of $(l,b)=(18.040\degr, -15.249\degr)$, an extinction of $E(B-V)=0.11\pm 0.018$ is expected from the 3D extinction maps based on Gaia parallaxes and stellar photometry \citep{Green2019}. The total extinction in the $G$-band is thus $A_{\mathrm{G}}=0.269 \pm 0.044$ \citep{Fitzpatrick1999, Schlafly2011, Wang_extinc_2019}, and the absolute magnitude results as $M_{\mathrm{G}}=2.01 \pm 0.2$ mag, which is consistent with the luminosity of sdO stars \citep{Geier2019}.

\begin{figure}
	\includegraphics[width=\columnwidth]{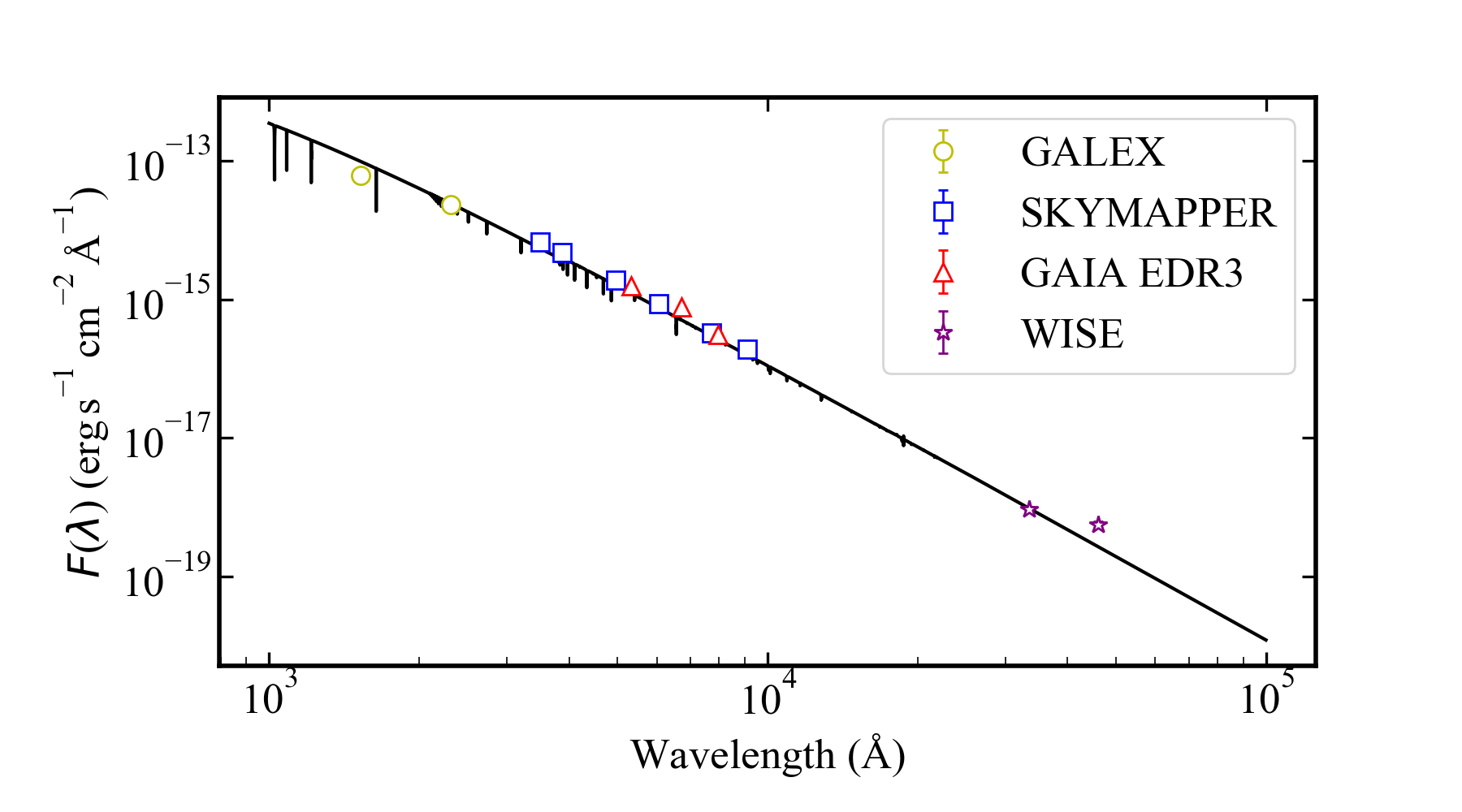}
    \caption{SED fitting of different bands. The yellow circles show the data from GALEX in $FUV$ and $NUV$ bands, the blue squares are the data from SkyMapper $uvgriz$ bands, the purple stars are the WISE $W1$ and $W2$ bands, and the red triangles are the data from Gaia eDR3 in the $BP$, $G$ and $RP$ bands. The black line is the best fitting model of the observational data using the TMAP model by SPEEDYFIT. }
    \label{fig:sed}
\end{figure}

We combined photometric data from the Galaxy Evolution Explorer (GALEX) \citep{GALEX2007}, Gaia EDR3 \citep{Gaia_edr3_2021}, SkyMapper \citep {SKYMAPPER2019} and Wide-field Infrared Survey Explorer (WISE) \citep{WISE2010} to analyse the spectral energy distribution (SED, see Figure~\ref{fig:sed}). We used the models of the T{\"u}bingen non-local thermodynamic equilibrium (NLTE) Model-Atmosphere package (TMAP) to fit the SED \citep{Werner2003}. An MCMC approach is used to find the minimum of the residuals between the observational SED and the TMAP model, and derive errors on the parameters. The code used is included in the SPEEDYFIT package \citep{Vos2012, Vos2013, Vos2017}.  We set the effective temperature and surface gravity as well as the distance as the prior assumptions taken from the spectral fitting and Gaia eDR3 distance \citep{Gaia_edr3_2021, Riess2021}. We find fitting parameters for the primary radius of $R_{\mathrm{sdO}}=0.47^{+0.30}_{-0.15}\,{\mathrm{R_{\odot}}}$ and a primary mass of $M_{\mathrm{sdO}}=0.55^{+0.94}_{-0.29}\,{\mathrm{M_{\odot}}}$, 
which are consistent with the measured mass $M_{\mathrm{sdO}}=0.55\,{\mathrm{M_{\odot}}}$ and radius $R_{\mathrm{sdO}}=0.47\,{\mathrm{R_{\odot}}}$ from the light curve fitting. 

\begin{table*}
  \centering
  \caption{Overview of the measured and derived parameters for J1920-2001.}
   \setlength{\tabcolsep}{0.7\tabcolsep}
  \begin{tabular}{lccc}
\hline\hline
Right ascension & RA [hrs] & 19:20:54.508 \\
Declination & Dec [deg] & -20:01:35.62 \\
Parallax & $\varpi$ [mas] & $0.14119\pm 0.03891$\\
Apparent magnitude & $m_{\mathrm{G}}$ [mag] & $16.53\pm0.009$\\
Absolute magnitude (reddening corrected) & $M_{\mathrm{G}}$ [mag] & $2.01\pm 0.2$\\
Proper motion in right ascension & $\mu_{\alpha}\cos({\delta})$ [mas $\mathrm{yr}^{-1}$] & $-4.364\pm0.072$\\
Proper motion in declination & $\mu_{\delta}$ [mas $\mathrm{yr}^{-1}$] & $-5.853\pm0.062$\\
Galactic radial velocity positive towards the Galactic centre & {\it U} [km/s] &282.79\\
Galactic rotational velocity in the direction of the Galactic & {\it V} [km/s] &86.99\\
Galactic velocity toward the North Galactic Pole & {\it W} [km/s] &7.14\\
Metallicity & {\it Z} & halo star\\
\hline
Atmospheric parameters of the sdO from spectra fitting& & \\
Effective temperature & $T_{\mathrm{eff}}$ [K] & 54\,240$\pm1\,840$ K\\
Surface gravity & $\log{g}$ & 4.841$\pm0.108$\\
Helium abundance & $\log{(n({\mathrm{He}})/n({\mathrm{H}}))}$ & $-1.190\pm0.142$\\
Projected rotational velocity & $v_{\mathrm{rot}}$sin$i$ [km/s] & $143.47\pm7.5$ \\
\hline
Orbital parameters from radial velocity curve fitting& &\\
Ephemeris zero point for 2015& $T_0$ [BJD] & 2\,457\,301.1279\\
Orbital period & $P_{\mathrm{orb}}$ [hour] & 3.4946\\
RV semi-amplitude (sdO) & $K$ [km/s] & $169.22\pm 2.03$\\
System velocity & $\gamma$ [km/s] & $200.95\pm 4.46$\\
Eccentricity & $e$ & 0\\
\hline
Derived parameters from light curve modelling& &\\
Mass ratio & $q=\frac{M_{\mathrm{WD}}}{M_{\mathrm{sdO}}}$ & $0.738\pm0.001$\\
sdO mass & $M_{\mathrm{sdO}}$ [$\mathrm{M_{\odot}}$] & $0.55$\\
sdO radius & $R_{\mathrm{sdO}}$ [$\mathrm{R_{\odot}}$] & $0.47$ \\
WD mass	 & $M_{\mathrm{WD}}$ [$\mathrm{M_{\odot}}$] & $0.41$\\
Orbital inclination & $i$ [deg] & $89.61^{+0.27}_{-0.44}$\\
\hline
Derived parameters from disk modelling & &\\
Disk radius &$R_{\mathrm{disk}}$ [$\mathrm{R_{\odot}}$]  & $0.41$\\
Disk thickness &$d$ [$\mathrm{R_{\odot}}$]  & $0.18$\\ 
\hline
Estimated parameters & &\\
WD radius &$R_{\mathrm{WD}}$ [$\mathrm{R_{\odot}}$]  & $0.0158$\\

\hline\hline
\end{tabular}
\label{tab:parameters}
\end{table*}

\subsection{The {\it U-V} velocity diagram}
\label{sec:UV}
\begin{figure}
	\includegraphics[width=\columnwidth]{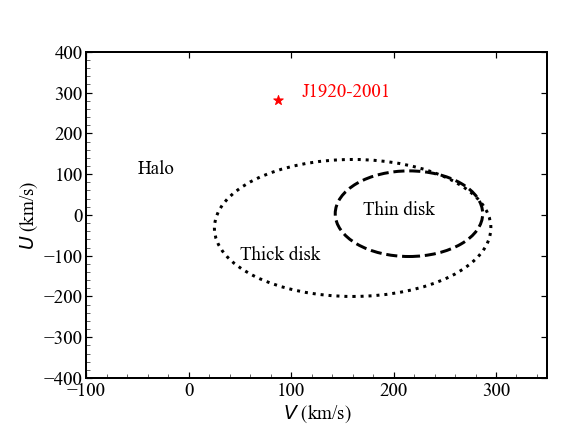}
    \caption{The {\it U-V} velocity diagram for J1920-2001 with thin disk and thick disk contours. The dashed line represents the 3$\sigma$ limits for the thin disk and the dotted line represents the thick disk within $3\sigma$ \citep{Pauli2006}. }
    \label{fig:UVdiagram}
\end{figure}

The {\it U-V} velocity diagram is a classical tool for kinematic investigations which is used to distinguish the populations of WD \citep{Pauli2006}. 
\cite{Luo2019} applied the {\it U-V} velocity diagram to distinguish the populations of hot subdwarfs. 
Adopting a parallax of $0.14119\pm 0.03891$ mas for J1920-2001 and its proper motion of PM(RA)$=-4.364\pm0.072$ mas\,$\mathrm{yr}^{-1}$ and PM(Dec)$=-5.853\pm0.062$ mas\,$\mathrm{yr}^{-1}$, we obtained the tangential velocities. Combining the tangential velocities with the radial velocity (200.95\,km/s), we obtained the space velocities
 (See also Sec~\ref{subsec:Orbital_period}).
In the Cartesian Galactic coordinates, {\it U} is 282.79\,km/s and {\it V} is 86.99\,km/s.
The location of J1920-2001 in the {\it U-V} diagram is shown in Figure~\ref{fig:UVdiagram}, which clearly shows that J1920-2001 could be a halo star and should have a correspondingly low metallicity and a large age.

\section{Formation Channels and Destiny}
\label{sec:formation}

The location of the sdO star of J1920-2001 in the $T_{\rm eff}$-$\log g$ diagram, or the Kiel diagram, implies that the sdO star has a mass of $\sim 0.58M_\odot$, a large CO core, and a thin burning shell(s). To produce such a binary, we can have two possible formation channels.
The first one is the canonical CE channel in which a star fills its Roche lobe in the red giant branch and experiences a CE phase. After the CE ejection, the helium core of the star is ignited and results in the formation of the sdO binary. The sdO binary evolves, burns helium into CO in its centre, until the sdO fills its Roche lobe and is observed as is the case here. Another, more likely, the channel is a CE ejection during the early asymptotic giant branch (AGB) phase. A star evolves to its early AGB phase, where it has a CO core surrounded by a burning shell(s) and a hydrogen envelope. The star fills its Roche lobe and then experiences a CE evolution. The CE ejection then leads to the formation of a tight binary. The remnant of the primary, the sdO star, has a CO core, surrounded by a burning shell(s). The matter outside of the inner Roche lobe is ejected during the CE phase, and the sdO binary is born as Roche-lobe filling. The matter inside the Roche lobe of the WD may be left as something similar to a disk. While such a channel has not been discussed before and is speculative, we attempt a qualitative analysis in this section.

To investigate the above two channels, we adopt the latest updated code Modules for Experiments in Stellar Astrophysics (MESA) version 12.21.1 \citep{Paxton2011,Paxton2013,Paxton2015,Paxton2018,Paxton2019} to construct binary evolution models. The metallicity is set to be $0.001$ as the star is fairly certainly a halo star (see Section~\ref{sec:UV}).

\subsection{RGB CE channel}
\label{subsubsec:rgb}
\begin{figure}
	\includegraphics[width=\columnwidth]{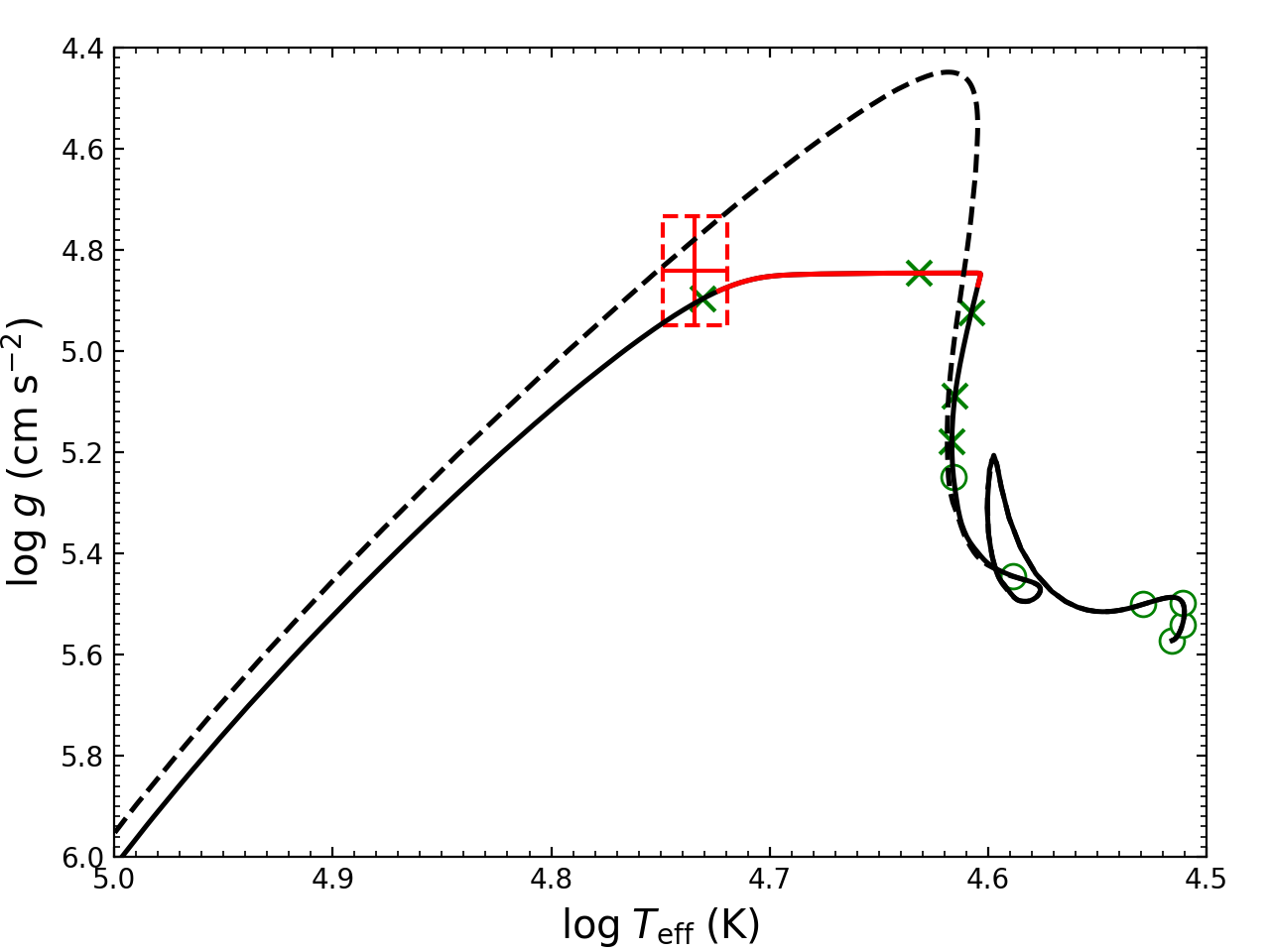}
    \caption{The evolutionary tracks of sdOs of $0.58\,M_\odot$ with a He-rich envelope of $0.05\,M_\odot$. The observed constraints are shown by the red box. The solid and dashed lines are for the evolution of a sdO binary and a single sdO star, respectively. The mass transfer phase is shown as the red solid line. The green circles and crosses mark time intervals of 10 Myr and 1 Myr, respectively. 
    }
    \label{fig:1}
\end{figure}

Given the rather large sdO mass and the rather tight orbital period, we need the progenitor of the sdO star to be massive enough and to experience a CE at the base of the RGB branch. 
We first construct the sdO star from a progenitor star with a mass of $3.8\,M_\odot$. As the star ascends the RGB and the He core mass grows to $0.53\,M_\odot$, we strip off the outer envelope until a He-rich envelope\footnote{The envelope mass is defined as the difference between surface mass and the He core mass, where the boundary between the envelope and the core is defined by a hydrogen mass fraction of $0.1$. According to the element fraction profile of the sdO star, we found that the envelope is enriched in He due to the nuclear burning (see also \citealt{Kupfer2020a,Bauer2021}), and the envelope comprises $70.9\%$ He and $25.9\%$ H.} of $\sim0.05\,M_\odot$ is left. We then evolve the sdO binary with a WD companion of $0.41\,M_\odot$ and an initial orbital period of 0.148\,d. The evolutionary tracks of sdOs are shown in Figure \ref{fig:1}, where the solid and the dashed lines are for the evolution of a sdO binary and a single sdO star, respectively. The evolution of the sdO stars is mainly driven by the core He burning (converting the helium core to a CO core) and the hydrogen shell burning, while the core He burning dominates the evolution of the sdO star before the onset of mass transfer. The timescale from the birth of sdO to the occurrence of mass transfer is about $6.3\times 10^{7}\,\rm yr$, and the mass of He-rich envelope decreases to $\sim 0.02\,M_\odot$ due to the residual hydrogen burning in the shell.  During the mass transfer phase, the accumulation efficiency of the WD companion is very low due to H or He shell flashes \citep{Kato2004}. For simplicity, we assume that all of the transferred mass is lost from the binary system and takes away the specific orbital angular momentum as pertains to the WD. The mass transfer lasts about $1\,\rm Myr$, and only $\sim0.007\,M_\odot$ envelope mass is lost from the donor. The final orbital period at the termination of mass transfer is about $1.4497\,\rm d$. Note that we did not do any fine-tuning to make the sdO mass to be exactly the same as that derived observationally. The mass is, however, very close to the observational one. For comparison, the evolution of a single sdO with a similar CO core mass and an envelope mass is shown as a dashed line.

The orbital period change during the mass transfer is affected by the mass transfer rate and orbital angular momentum loss, and can be calculated as \citep{postnov2014}.
\begin{eqnarray}
  \frac{\dot{P}_{\rm orb}}{P_{\rm orb}} = \frac{\dot{M}_{\rm sdO} }{M_{\rm sdO} }\left(\frac{M_{\rm sdO}}{M}-3\right) + \frac{3\dot{J}_{\rm orb}}{J_{\rm orb}},
\label{eq:x1}
\end{eqnarray}
where $M = M_{\rm sdO}+M_{\rm WD}$, and $J_{\rm orb}$ is orbital angular momentum. In our simulations, the maximum mass transfer rate is $10^{-8.2}{M_\odot\,\rm yr^{-1}}$, so the maximum period derivative can be obtained with equation~(\ref{eq:x1}) as $0.0007 \rm \,s\,yr^{-1}$. This value is significantly smaller than the observed value of $0.017 \,\rm s\,\rm yr^{-1}$, or 0.1\,s in 6 yr. In order to solve this contradiction, we consider a more efficient angular momentum loss mechanism. We now assume that the WD has a high magnetic field and that mass is lost from the WD magnetosphere. The magnetosphere radius of the magnetic WD is calculated as \citep{hameury1987}. 
\begin{eqnarray}
  R_{\rm mag} &=& 4.13\times10^{14} \left(\frac{B_{\rm mag}}{10^{9}\,\rm G}\right)^{4/7}\left(\frac{R_{\rm WD}}{R_{\odot}}\right)^{12/7} \\ \nonumber
  && \left(\frac{\dot{M}}{10^{-8}M_\odot\,\rm yr^{-1}}\right)M_{\rm WD}^{-1/7}, 
  \label{eq:x2}
\end{eqnarray}
where $B_{\rm mag}$ is the surface magnetic field of WD. For an extreme case of $B_{\rm mag}=10^9\,\rm G$ \citep{ferrario2015}, the magnetosphere radius is about $426\,R_\odot$. Assuming that the lost mass carries away the specific orbital angular momentum at the magnetosphere, the period change is calculated to be $0.02\,\rm s\,\rm yr^{-1}$ according to equation~(\ref{eq:x1}). 
However, there is no evidence for such a strong magnetic field in the observations that may support the period change of $0.02\,\rm s\,\rm yr^{-1}$.
The dramatic orbital shrinkage in the observation may suggest that there are some other efficient angular momentum loss mechanisms, such as the friction of circumbinary disk \citep{spruit2001,willems2005}. The AGB CE channel presented in the next Section may provide just such a friction.

\subsection{AGB CE channel}
\label{subsec:AGB}
\begin{figure}
	\includegraphics[width=\columnwidth]{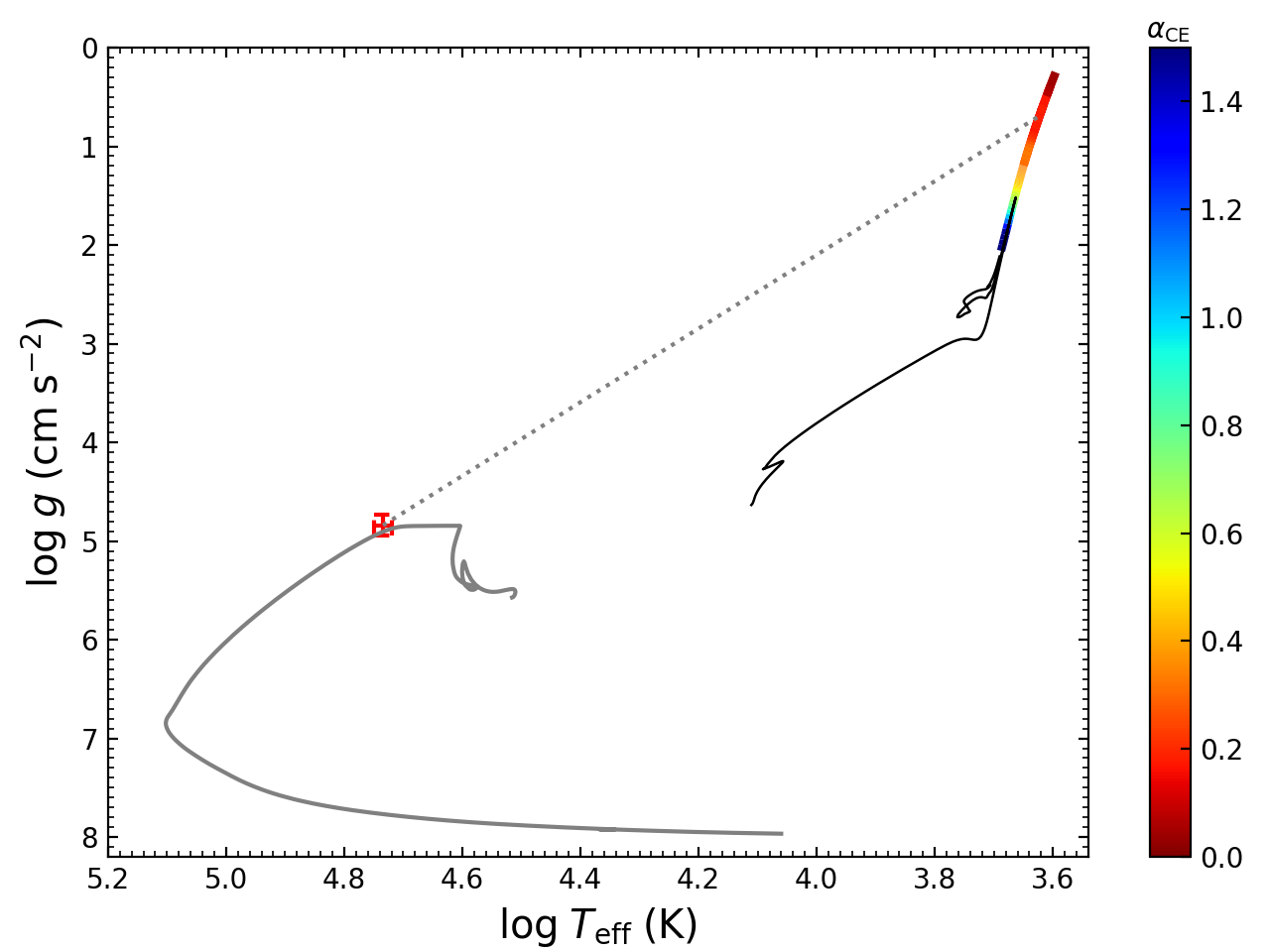}
    \caption{The values of $\alpha_{\rm CE}$ for a $2.1\,{\mathrm{M_\odot}}$ progenitor at AGB stage. The evolutionary track for the star from MS to AGB is shown in the black line, and the values of $\alpha_{\rm CE}$ are shown in colours. The dotted line represents the evolution direction during the CE process. The constructed model of sdO is taken from Section~\ref{subsubsec:rgb}, as shown in the grey line. }
    \label{fig:3}
\end{figure}

An early AGB star has a quite large CO core and a helium (hydrogen) burning shell(s) and a hydrogen envelope,  which is appropriate for the progenitor of the sdO star of J1920-2001. 
In this channel, the star fills its Roche lobe at the early AGB stage, and the mass transfer rate is so high that a CE forms. Next, a sdO binary can be produced if the CE is ejected. The sdO binary produced this way would have a CO core and a burning shell(s). To see whether the CE can be ejected or not, we adopt the energy prescription: if the orbital energy released during the spiral-in process overpowers the binding energy of the CE, the CE is assumed to be ejected \citep{livio1988,dekool1990}. We first evolve a $2.1\,\rm M_\odot$ star from MS to AGB and try to produce a sdO star with a mass of $0.58\,\rm M_\odot$ (as constructed in Section~\ref{subsubsec:rgb}, via CE ejection). The WD companion of the AGB star has a mass of $0.41\,\rm M_\odot$, and the orbital period after CE ejection is set to be $0.148\,\rm d$. Then the CE efficiency, $\alpha_{\rm CE}$, where the star fills the Roche lobe at the AGB stage, can be calculated as \citep{lizw2019}
\begin{eqnarray}
  \alpha_{\rm CE} = \frac{E_{\rm bind}}{G(M_{\rm sdO}+M_{\rm env})M_{\rm WD}/(2a_{\rm i})-GM_{\rm sdO}M_{\rm WD}/(2a_{\rm f})},
\end{eqnarray}
where $E_{\rm bind}$ is the binding energy of the envelope, $M_{\rm env} = 2.1\,{\rm M_\odot} - M_{\rm sdO}$ is the envelope mass, $a_{\rm i}$ is the initial orbital separation and is set to make the AGB star fill its Roche lobe, and $a_{\rm f}$ is the final orbital separation and is set for the final orbital period of 0.148\,d. The values of $\alpha_{\rm CE}$ during the AGB stage are shown in Figure~\ref{fig:3}. In most cases, the CE efficiencies $\alpha_{\rm CE}$ are smaller than $1$, indicating that the CE may be ejected successfully at the AGB stage. The evolution of the sdO star during the CE phase is rather unclear, so we use a straight dotted line to sketch its direction.  
The subsequent evolution of the sdO star is similar to that in Section~\ref{subsubsec:rgb} and shown in the grey line. 

The sdO binary produced in this way is a Roche lobe filling binary at birth. Some of the CE mass has not dispersed completely and could produce the friction needed for the drastic orbital shrinkage which cannot be explained by mass loss and conventional angular momentum loss alone. The orbital shrinkage would be less severe after the CE mass has dispersed.

To have an idea of how much density of the CE remnant material is needed to produce the drag, we have made an estimate as follows. The sdO star has a much larger radius than the WD and we assume that the sdO star interacts with the CE remnant material to produce the drag. We assume that the sdO star expels the CE mass with a velocity $V_{\rm sdO}$ and a cross-section of $\pi R_{\rm sdO}^2$, where $V_{\rm sdO}$ and $R_{\rm sdO}$ are the orbital velocity and the radius of the sdO star, respectively. 
The orbital angular momentum loss, $\Delta J_{\rm{orb}}$, is written as,
\begin{equation}
    \Delta J_{\rm{orb}}=\pi R_{\rm sdO}^2 V_{\rm sdO} \Delta t \rho V_{\rm sdO} {M_{\rm WD}\over M_{\rm sdO}+M_{\rm WD}}A
    \label{eq:dj}
\end{equation}
where $\Delta t$ is the time interval (i.e. 6 yr in this study) for the angular momentum loss, $\rho$ the density of the remnant CE material, and $A$ the orbital separation of the binary system.
We further rewrite the above equation as,
\begin{equation}
{\Delta J_{\rm{orb}}\over J_{\rm{orb}}}={3\over 2}\pi ({q\over 1+q})^2({\Delta t\over P_{\rm{orb}}}) ({A\over R_{\rm sdO}}) ({\rho\over \rho_{\rm sdO}})
\label{eq:djj}
\end{equation}
where $J_{\rm{orb}}$ is the orbital angular momentum of the binary system, $q={M_{\rm WD}\over M_{\rm sdO}}$, $P_{\rm{orb}}$ is the orbital period, and $\rho_{\rm sdO}={M_{\rm sdO}\over {4\over 3}\pi R_{\rm sdO}^3}$ is the average density of the sdO star.
The relation between the orbital period change and the orbital angular momentum loss can be written as,  
\begin{equation}
    {\Delta J_{\rm{orb}} \over J_{\rm{orb}}} ={1\over 3} {\Delta P_{\rm{orb}}\over P_{\rm{orb}}}
    \label{eq:djdp}
\end{equation}
Combining equations ~\ref{eq:djj} and ~\ref{eq:djdp}, we have
\begin{equation}
    \rho={2\over 9\pi} (1+{1\over q})^2{R_{\rm sdO}\over A}{\Delta P_{\rm{orb}}\over \Delta t}\rho_{\rm sdO}
    \label{eq:rho}
\end{equation}
The sdO star is filling its Roche lobe, and we adopt Eggleton's Roche lobe radius equation \citep{Eggleton1983}, 
\begin{equation}
    {R_{\rm L, sdO}\over A}={0.49q_{\rm sdO}^{2/3}\over 0.6q_{\rm sdO}^{2/3}+\ln{(1+q_{\rm sdO}^{1/3})}}
    \label{eq:roche}
\end{equation}
where $R_{\rm L, sdO}$ is the Roche lobe radius of the sdO star, and $q_{\rm sdO}={M_{\rm sdO}\over M_{\rm WD}}$. From equation ~\ref{eq:roche} we obtain the value of 
${R_{\rm sdO}}\over A$ to be 0.40. Substituting this value for $R_{\rm sdO}\over A$ in equation ~\ref{eq:rho}, we obtain the remnant CE density required for the drag to be $\rho\sim 8\times 10^{-11}\rho_{\rm sdO}$, or $\sim 6\times 10^{-10} {\rm g/cm^3}$.

Note that, in this channel, the progenitor binary system may have experienced a stable Roche lobe overflow to produce a WD + MS system first. Later the MS star evolves to early AGB phase and has an unstable mass transfer leading to the formation of a CE. The CE ejection may leave a system configuration like this: a sdO + a WD + some remnant CE materials sitting around. We speculate that the CE ejection removes the material outside of the Roche lobe of the sdO star and that of the WD. Consequently, the sdO star is Roche-lobe filling at birth (CE ejection) and the CE materials remained in the Roche-lobe of the WD condensed to the equatorial plane to form a disk as the materials have enough angular momentum. The Roche-lobe filling sdO star transfers mass to the WD and the disk is maintained this way.

\subsection{Destiny of the binary}
\label{subsec:Destiny}
\begin{figure}
	\includegraphics[width=\columnwidth]{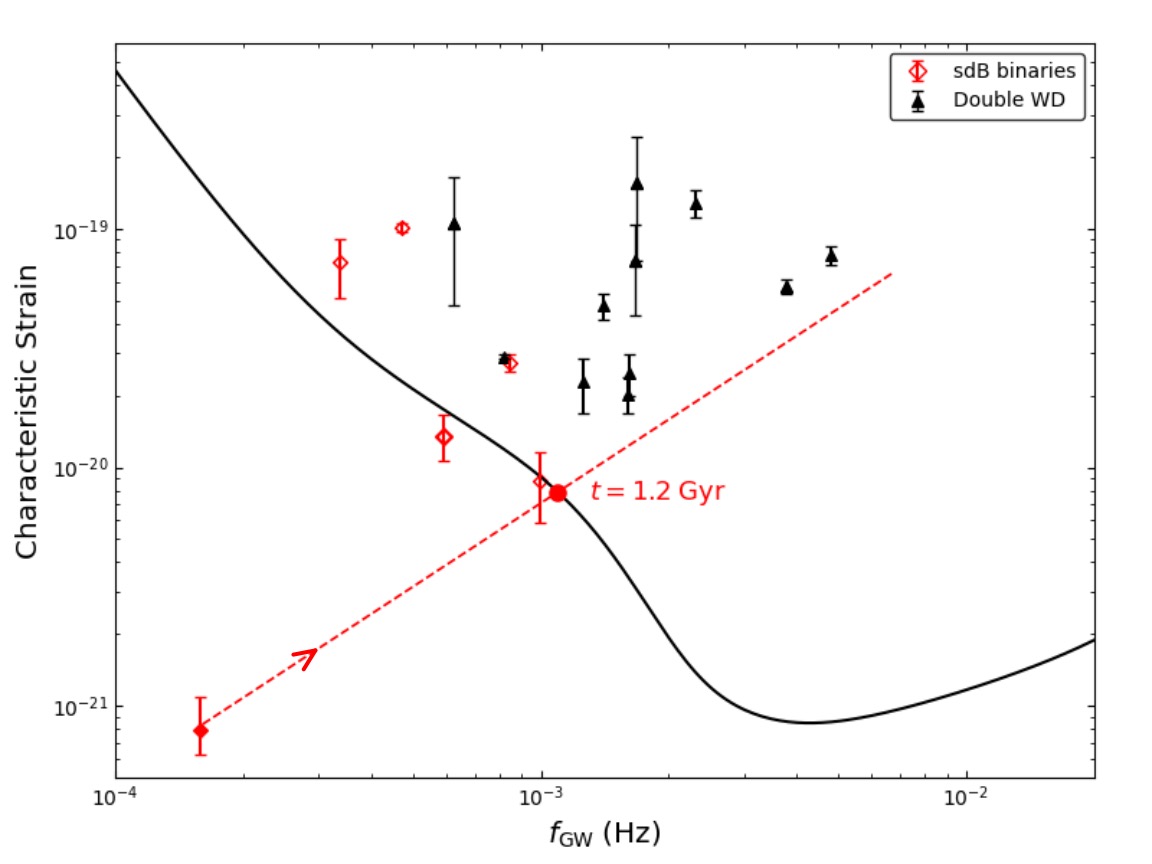}
    \caption{The characteristic strain versus gravitational-wave frequency for observed samples. The verification sources of double WDs and sdB binaries are shown in black solid triangles and red open diamonds, respectively \citep{Geier2013,Kupfer2018,Kupfer2020a,Kupfer2020b,Burdge2019a,Burdge2019b,Burdge2020a,Burdge2020b,Pelisoli2021}. The red solid diamond is J1920-2001, and its future evolution is shown in the red dashed line. The black solid line is the LISA 4 yr sensitivity curve \citep{Robson2019}, and we see that the system may be detectable by a LISA-like gravitational wave radiation detector after $\sim 1.2\, \rm Gyr$ (the red filled circle). 
    }
    \label{fig:2}
\end{figure}

After the termination of the mass transfer, the sdO star continues to contract and the evolutionary tracks are similar to that of single sdO evolution. About $1$ Myr after the termination of mass transfer, the helium core is almost exhausted and the residual hydrogen burning also ceases. With the cooling of the core and the envelope, the sdO evolves to become a WD. The separation of the double WD binary will decrease due to angular momentum loss via gravitational-wave radiation. The relation between the characteristic strain and the gravitational wave frequency after the mass transfer phase is shown in Figure~\ref{fig:2}. About 1.2 $\rm Gyr$ after the end of mass transfer, the gravitational wave radiation of this binary may be detected by the Laser Interferometer Space Antenna (LISA, \citealt{lisa2017}), this stage is shown as a red point in Figure~\ref{fig:2}. The binary will come into contact again due to the decrease in orbital separation. The mass transfer is expected to be unstable in the double WD due to the large mass ratio \citep{Han1999, Marsh2004}, and the merger product of the double WD may be a single R CrB star \citep{Guillochon2010,Zhangx2014,Brown2016}.

\section{Ca~H\&K from the ejected common envelope?}
\label{sec:CaII}
\begin{figure}
	\includegraphics[width=\columnwidth]{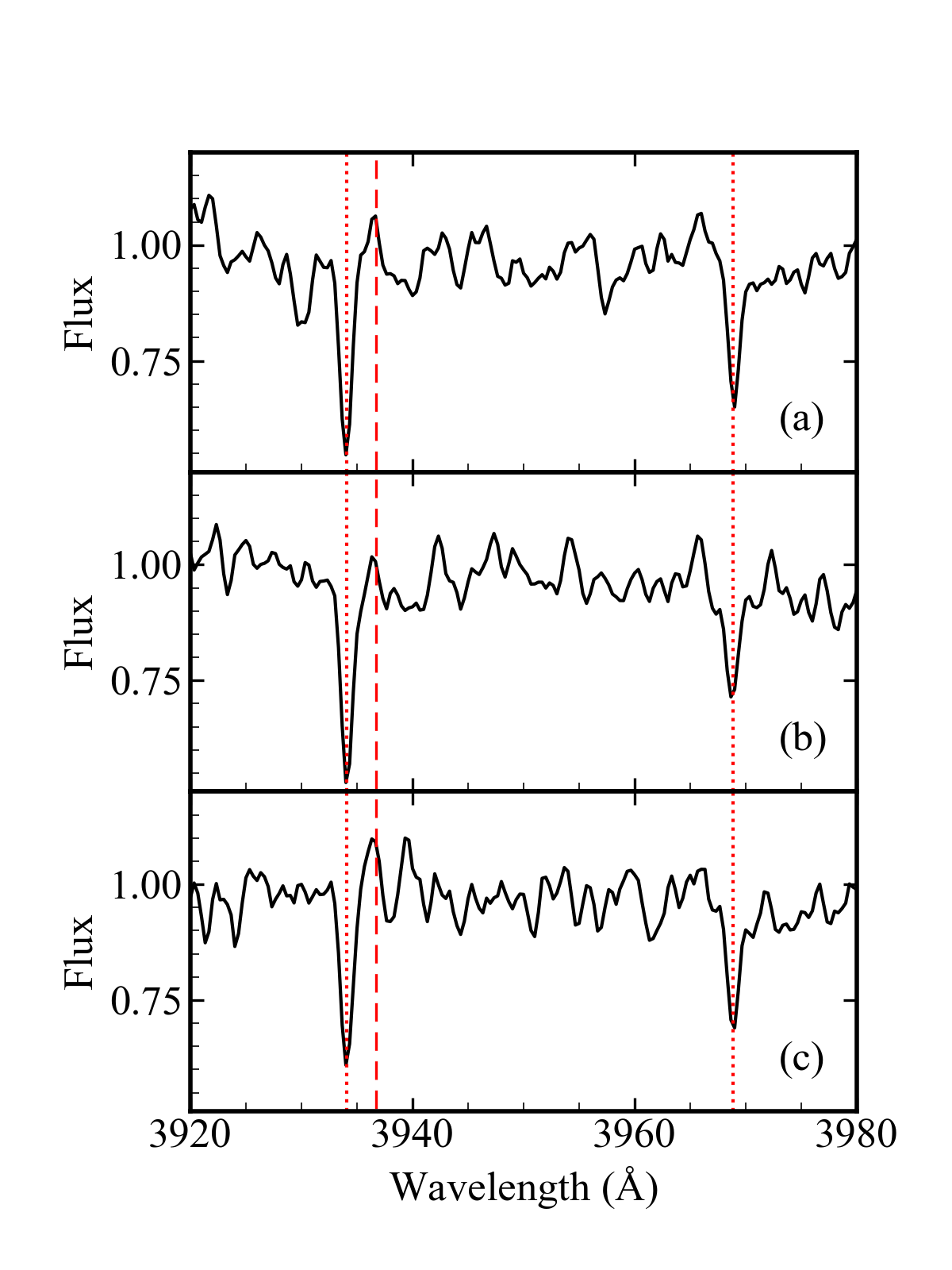}
    \caption{The profile of the Ca~II H\&K lines in three selected spectra from 25 June 2019 to 23 Oct 2019. Dotted lines label the Ca~II H\&K lines with a velocity of 29.93\,km/s relative to their rest wavelength. Dashed lines label the position of the systemic velocity of 200.95\,km/s redshifted to the Ca~II absorption lines. Hence, the observed absorption lines have a blueshift of 171\,km/s relative to the mass centre of the binary. Near the velocity of the mass centre (dashed line) a tentative emission line is seen to produce a P Cygni profile, which can originate from an expanding shell. The Ca~H line is blended with the H$\epsilon$ Balmer line so that the emission part is not visible.
    }
    \label{fig:CaII}
\end{figure}

An outstanding feature in the observed spectra are stationary, strong, and sharp lines of Ca~II H\&K (see Figure~\ref{fig:CaII}), which are clearly not associated with the photosphere of a hot star, but must arise from an intervening medium. We show here that the absorption lines are almost certainly inconsistent with the reddening caused by the interstellar medium (ISM) and are hence more likely caused by circumstellar medium (CSM).

\begin{figure}
	\includegraphics[width=\columnwidth]{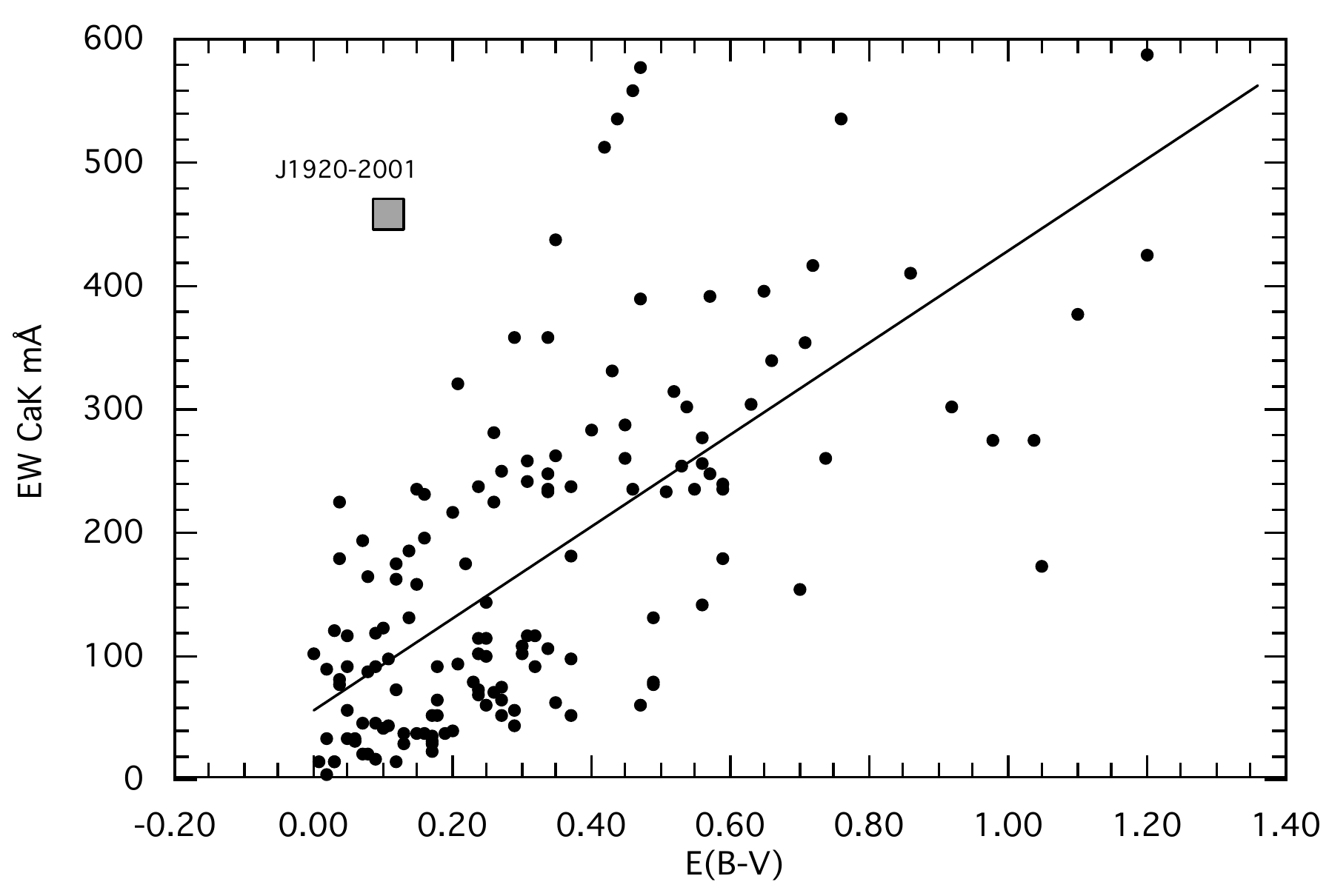}
    \caption{The mean ISM relation between the Ca~K line strength and $E(B-V)$ for OB stars compared to that for J1920-2001. The dots are data from \citet{Megier2005} for a large sample of OB stars at different lines of sight through the interstellar medium and the solid diagonal line is a linear fit. Our data point, shown as a grey square symbol, is located far above the line.
    }
    \label{fig:CaK}
\end{figure}

Figure~\ref{fig:CaK} shows a relation between the Ca~K line intensity and the $E(B-V)$ reddening.
In the ISM, the mean relationship between the Ca~II~K (3\,933\,\AA) line strength and the $E(B-V)$ for field stars \citep{Megier2005} suggests that a K line strength of about 460\,m\AA, as observed in J1920-2001, corresponds to a mean $E(B-V)\approx 1$\,mag, while the value suggested by reddening maps was only $\sim$0.1\,mag. A Ca~K line that is $\sim5$ times stronger than predicted by interstellar extinction must therefore originate from a gas-rich but dust-poor region associated with the hot star. 

Figure~\ref{fig:CaII} shows that the Ca~H\&K absorption lines are blueshifted by $\sim$171\,km/s relative to the mass centre of the binary. The lines show a marginal P~Cygni profile, with the marginal emission lines at zero velocity relative to the mass centre. Irrespective of the tentative emission component, the absorption components suggest that the lines arise from a circumbinary shell expanding with a velocity of $\sim$171\,km/s. 

For the RGB CE channel mentioned above, we assume that the progenitor of the sdO is a star at the base of the red giant branch with a mass of $3.8\,{\mathrm{M_\odot}}$. The mass of the ejected CE is around $3.2\,{\mathrm{M_\odot}}$, and the mass of the system is $M_{\mathrm{sdO}}+M_{\mathrm{WD}}+M_{\mathrm{CE}} \sim4.2\,{\mathrm{M_\odot}}$. The radius at the base of the RGB for a $3.8\,{\mathrm{M_\odot}}$ star is $32\,{\mathrm{R_\odot}}$, and the binary separation at the onset of CE is around $64\,{\mathrm{R_\odot}}$, which we take as the radius of the CE. By assuming that the ejection velocity must be larger than the escape velocity of the CE, with $v_{\mathrm{esc}}=\sqrt{2GM/R}=158.2$\,km/s, we find the shell expansion velocity derived from the Ca~H\&K lines to be close to but indeed larger than $v_{\mathrm{esc}}$. 

Similarly, we have an escape velocity of 69\,km/s of the CE for the AGB model. The AGB star has a larger radius, and consequently, this velocity is smaller than the shell expanding velocity derived. However, \citet{Clayton2017} carried out 1D hydrodynamic simulations for CE ejection, and the ejection is found to be episodic, i.e. ejected shell by shell. During the last shell ejection, we can speculate that the binary and the CE are much tighter and the escape velocity would then be higher. The ejection velocity can therefore possibly match the observed expansion velocity of the shell.

We now estimate the shell density from the observed Ca~II K line strength of $\sim$0.46\,\AA. We simply assume that the line is of negligible saturation, and the absorption is then in the linear regime of the curve of growth. The column density $N({\rm Ca\ K})$ can be calculated from the line width $EW$ \citep{Ellison2004, Murga2015},  
\begin{eqnarray}
N({\rm Ca\ {K}})=1.13\times 10^{20}\, {\rm cm}^{-2} {\frac{EW}{f \lambda ^2}},
\end{eqnarray} 
where both $EW$ and $\lambda$ are in unit of {\AA}. The oscillator strength $f$ for the Ca~II K line is 0.6485 \citep{Safronova2011}, and we therefore obtain the Ca~K column density to be $N({\rm Ca\ K})=5.2\times 10^{12}\, {\rm cm}^{-2}$.  Neglecting depletion by Ca element condensation and taking $\log(N({\rm Ca\ II})/N({\rm H}))=-5.67$ to be the same as the solar photospheric abundance, we have $N({\rm H})=2.4\times 10^{18}\, {\rm cm}^{-2}$. Such a hydrogen number density corresponds to a column density of $5.7\times 10^{-6}\, {\rm g}\, {\rm cm}^{-2}$ for a shell with a hydrogen mass fraction of $X=0.7$. 

For the RGB CE channel, the time from the birth of the sdO star to the current state, i.e. the time elapsed since the CE ejection, is about $6\times 10^7$ yrs (as shown in Figure~\ref{fig:1}).  Assuming the CE with a mass of $3.2\,M_\odot$ is ejected at a constant speed of 158.2\,km/s and the material is distributed evenly in the expanding spherical shell, the column density is around $5.7\times10^{-13}$\,$\mathrm{g\, cm^{-2}}$, which is lower than that required by 7 orders of magnitude. The discrepancy would be bigger if any line saturation or element condensation exists.

For the AGB CE channel, the ejected mass during the CE process is $1.52\,M_\odot$, and we need a spherical surface of $5.3\times 10^{38}\,{\rm cm}^2$ to get the required column density. The surface corresponds to a radius of $6.5\times 10^{18}\,{\rm cm}$. We, therefore, need an ejection age of 
$\sim 10\,000\,{\rm yrs}$ for an ejection velocity of 171\,km/s. The ejection age would be smaller if line saturation or element condensation exists to some extent. The time scale of CE ejection is a few thousand yrs (e.g. \citet{Clayton2017}), and we may be observing a CE ejected quite recently.

We did not do any detailed analysis of the H\&K lines. The analysis here is rather qualitative. However, we can see that it is very likely for lines to originate from a CE ejected very recently. More detailed follow-up observations and simulations of the formation of the H\&K lines may provide a better understanding of the origin of the blueshifted Ca~II lines.


\section{Summary}
\label{sec:summary}
J1920-2001 was discovered as a sdO+WD binary with an orbital period of $P_{\mathrm{orb}}=0.145609$\,d (approx. 3.5 hours). The shape of the phased light curve is similar to that in the recently discovered binaries ZTF J2130+4420 and ZTF J2055+4651. Photometric data from {\it K2} in combination with spectroscopy obtained with WiFeS constrain the system parameters to a sdO star with an effective temperature of $T_{\mathrm{eff}}=54\,240\pm1\,840$\,K and a surface gravity of $\log{g}=4.841\pm0.108$, a mass of $M_{\mathrm{sdO}}=0.552\,{\mathrm{M_{\odot}}}$, an inclination angle of the binary orbit of $89.61^{+0.27}_{-0.44}$ degree, a mass ratio of $q=M_{\mathrm{WD}}/M_{\mathrm{sdO}}=0.738\pm 0.001$.
The disk modelling shows the disk radius of $\sim 0.41$\,$R_{\odot}$ and the disk thickness of $\sim 0.18$\,$R_{\odot}$.
The parameters of the SED fitting are consistent with the parameters above.

We proposed two channels for the formation of the binary, a CE ejection at the RGB stage and a CE ejection at the early AGB stage. The AGB CE ejection channel is more likely: it can easily explain the shrinkage of the orbital period of 0.1\,s in 6 yr and the blueshifted Ca~II H\&K lines. The shrinkage can be due to friction between the binary and the remnant CE material in the binary system (with a CE remnant mass density of $\sim 6\times 10^{-10} {\rm g /cm^3}$), and the Ca~II H\&K lines are probably from a very recently ejected CE. Further observations and simulations are needed to get a better understanding of the orbital shrinkage and the Ca~II H\&K lines. This could help greatly address the long-standing problem of CE evolution, which is the most important and the least understood process in binary evolution.

\section*{Acknowledgements}
We thank Philipp Podsiadlowski for discussions and the anonymous referee for his/her valuable comments. This work is supported by National Natural Science Foundation of China (Grant Nos. 12090040/3, 12125303, 11733008, 12103086), National Key R$\&$D Program of China (Gant No. 2021YFA1600401/3). We also acknowledge the science research grant from the China Manned Space Project with No.CMS-CSST-2021-A10.
CAO was supported by the Australian Research Council (ARC) through Discovery Project DP190100252.
PN acknowledges support from the Grant Agency of the Czech Republic 
(GA\v{C}R 22-34467S) and from the Polish National Science Centre under projects No.\,UMO-2017/26/E/ST9/00703 and UMO-2017/25/B/ST9/02218.
The Astronomical Institute in Ond\v{r}ejov is supported by the project RVO:67985815.
We thank Thomas Behrendt from University of Melbourne for observations of the object with the ANU 2.3m-telescope in 2021.
This work has made use of data from the European Space Agency (ESA) mission
{\it Gaia} (\url{https://www.cosmos.esa.int/gaia}), processed by the {\it Gaia}
Data Processing and Analysis Consortium (DPAC,
\url{https://www.cosmos.esa.int/web/gaia/dpac/consortium}). Funding for the DPAC
has been provided by national institutions, in particular the institutions
participating in the {\it Gaia} Multilateral Agreement. This publication makes use of data products from the Wide-field Infrared Survey Explorer, which is a joint project of the University of California, Los Angeles, and the Jet Propulsion Laboratory/California Institute of Technology, funded by the National Aeronautics and Space Administration.
This paper includes data collected by the {\it Kepler} mission and obtained from the MAST data archive at the Space Telescope Science Institute (STScI). Funding for the {\it Kepler} mission is provided by the NASA Science Mission Directorate. STScI is operated by the Association of Universities for Research in Astronomy, Inc., under NASA contract NAS 5–26555.
The national facility capability for SkyMapper has been funded through ARC LIEF grant LE130100104 from the Australian Research Council, awarded to the University of Sydney, the Australian National University, Swinburne University of Technology, the University of Queensland, the University of Western Australia, the University of Melbourne, Curtin University of Technology, Monash University and the Australian Astronomical Observatory. SkyMapper is owned and operated by The Australian National University's Research School of Astronomy and Astrophysics. The survey data were processed and provided by the SkyMapper Team at ANU. The SkyMapper node of the All-Sky Virtual Observatory (ASVO) is hosted at the National Computational Infrastructure (NCI). Development and support of the SkyMapper node of the ASVO have been funded in part by Astronomy Australia Limited (AAL) and the Australian Government through the Commonwealth's Education Investment Fund (EIF) and National Collaborative Research Infrastructure Strategy (NCRIS), particularly the National eResearch Collaboration Tools and Resources (NeCTAR) and the Australian National Data Service Projects (ANDS).
This research made use of Lightkurve, a Python package for {\it Kepler} and TESS data analysis \citep{LightkurveCollaboration2018}.
This research has used the services of \mbox{\url{www.Astroserver.org}} under reference B8LQ7J.
\section*{Data Availability}

The photometric data underlying this article is available in the public archives of K2 and ZTF. Photometric light curves from SkyMapper and spectroscopic data will be shared upon reasonable request to the corresponding author.



\bibliographystyle{mnras}
\bibliography{ms} 


\

\bsp	
\label{lastpage}
\end{document}